\documentclass[twocolumn,preprintnumbers,superscriptaddress,prd,amsmath,amssymb,nofootinbib]{revtex4}
\usepackage{graphicx}
\usepackage{amssymb,amsmath}

\newcommand{\Neff}{N^{\rm eff}_\nu}
\newcommand{\TnuSM}{T^{\rm SM}_\nu}
\newcommand{\NnuSM}{N^{\rm SM}_\nu}
\newcommand{\Nnuint}{N^{\rm int}_\nu}
\newcommand{\rn}{{\rm n}}

\begin{document}

\preprint{KRL-MAP-309}

\title{Cosmological Signatures of Interacting Neutrinos}

\author{Nicole F. Bell}
\email{nfb@caltech.edu}
\affiliation{Kellogg Radiation Laboratory, California Institute of Technology, 
Pasadena, CA 91125}
\affiliation{Theoretical Astrophysics, California Institute of Technology, 
Pasadena, CA 91125}
\author{Elena Pierpaoli}
\email{pierpa@caltech.edu}
\affiliation{Theoretical Astrophysics, California Institute of Technology, 
Pasadena, CA 91125}
\author{Kris Sigurdson}\altaffiliation{Hubble Fellow}
\email{krs@ias.edu}
\affiliation{Theoretical Astrophysics, California Institute of Technology, 
Pasadena, CA 91125}
\affiliation{School of Natural Sciences, Institute for Advanced Study, 
Princeton, NJ 08540}

\date{March 29, 2006}

\begin{abstract}
  We investigate signatures of neutrino scattering in the Cosmic
  Microwave Background (CMB) and matter power spectra, and the extent
  to which present cosmological data can distinguish between a free
  streaming or tightly coupled fluid of neutrinos.  If neutrinos have
  strong non-standard interactions, for example, through the coupling
  of neutrinos to a light boson, they may be kept in equilibrium until
  late times.  We show how the power spectra for these models differ
  from more conventional neutrino scenarios, and use CMB and large
  scale structure data to constrain these models.  CMB polarization
  data improves the constraints on the number of massless neutrinos,
  while the Lyman--$\alpha$ power spectrum improves the limits on the
  neutrino mass.  Neutrino mass limits depend strongly on whether some
  or all of the neutrino species interact and annihilate.  The present
  data can accommodate a number of tightly-coupled relativistic
  degrees of freedom, and none of the interacting-neutrino scenarios
  considered are ruled out by current data --- although considerations
  regarding the age of the Universe disfavor a model with three
  annihilating neutrinos with very large neutrino masses.

\end{abstract}


\maketitle


\section{Introduction}
 
We are in a remarkable era when cosmological data is both precise and
abundant.  Measurements of temperature and polarization fluctuations
in the cosmic microwave background (CMB) radiation
(e.g. \cite{WMAP,ACBAR,CBI,CBIpol}), the matter power spectrum via
galaxy surveys \cite{2dF,SDSS} and the Lyman-$\alpha$ forest
\cite{VHS04,lyman}, the current and past expansion rate of the
Universe via the Hubble Key Project \cite{HST} and observations of
Type Ia supernovae \cite{Supernovae} respectively, and the abundance
of light elements predicted by big bang nucleosynthesis (BBN)
\cite{BBN} all show striking consistency within the standard
$\rm{\Lambda CDM}$ cosmological model.  Robust bounds can be placed on
the forms of matter and energy that constitute the Universe, and
cosmology is now a powerful particle physics laboratory that
constrains the properties of both the new dark-matter particles
demanded by cosmological data (e.g. \cite{DarkMatterProperties}) and
the familiar particles of the standard model of particle physics.  In
many cases these constraints would be difficult or impossible to
obtain in any other way.

Neutrino physics is an excellent example.  Cosmological techniques
for probing neutrino properties rely upon detecting indirect
signatures of the relic neutrino background, which complements the
significant experimental and theoretical effort that is underway to
understand the surprising physics of the neutrino sector.  A
particular focus of this effort has been to set limits on the mass and
number density of relic neutrinos.  In this work we instead focus on
the cosmological signatures of neutrino \emph{interactions}.

It is well known that the CMB can be used to constrain the number of
light relativistic degrees of freedom (in addition to the photon) --
conventionally parametrized as the effective number of neutrino
species, $N_\nu^{\rm eff}$.  A relative increase in the radiation density
delays the epoch of matter-radiation equality, and leads to an
enhanced Integrated Sachs-Wolfe (ISW) effect.  Present CMB limits are
$1.6 \leq N_\nu^{\rm eff} \leq 7.1$~\cite{Nnu,Pierpaoli}.  Big Bang
nucleosynthesis (BBN) also constrains $N_\nu^{\rm eff}$, as additional
radiation increases the expansion rate and alters the expected
primordial abundance of helium.  The current BBN bound is (up to
various flavor dependent subtleties) $N_\nu^{\rm
eff}<3.3-4$~\cite{BBN}.

The standard cosmological model predicts $N_\nu^{\rm eff}\approx3.04$,
consisting of the three known neutrino species, plus a small
correction that accounts for the neutrino heating from
electron-positron annihilation and finite-temperature QED effects
(e.g. \cite{Lopez:1998aq}).  With only the three standard neutrino
species, BBN constraints, combined with neutrino mixing, no longer
permit the possibility of a significantly enhanced $N_\nu^{\rm eff}$
due to large chemical potentials~\cite{degen}.  However, it is
important to bear in mind that $N_\nu^{\rm eff}$ may include not only
neutrinos, but any light particles that are thermally populated in the
early universe.  In models containing sterile neutrinos or other light
relativistic degrees of freedom, the BBN limits may be substantially
modified, and $N_\nu^{\rm eff} > 3$ is still possible~\cite{kneller}.
(In addition, $N_\nu^{\rm eff} < 3$ can be obtained if the reheating
temperature following inflation is low~\cite{lowreheat}.)  Moreover,
the relativistic energy density may evolve between the time of BBN and
CMB decoupling, so $N^{\rm eff}_{\nu}|_{\rm BBN}$ and $N^{\rm
eff}_{\nu}|_{\rm CMB}$ need not be the same quantity
(e.g.~\cite{chacko,neutrinoless}).

In addition to the total relativistic energy density, cosmology can
also be used to probe interactions in the neutrino sector.  Neutrinos
with only standard model couplings interact via the weak force, and
decouple when $T \simeq 1$ MeV.  At later times, they free-stream, and
interact only via gravity.  In this paper we investigate a class of
models which feature extra, non-standard, neutrino interactions. In
these models, neutrinos interact strongly with a new scalar boson,
which is brought into thermal equilibrium though its coupling to the
neutrinos.  Rather than free-streaming, the neutrinos form a tightly
coupled fluid with the new scalar.

These models generically have non-standard values for $N_\nu^{\rm
eff}$, but perhaps more interestingly, the absence of neutrino
free-streaming leaves a distinctive signature in the CMB.  If the
neutrinos are part of a tightly coupled fluid, they are fully
characterized by density and velocity perturbations, and 
anisotropic stress is negligible.  In~\cite{hannestad,trotta} it was
shown that the current Wilkinson Microwave Anisotropy Probe (WMAP) CMB
measurements already have some sensitivity to this effect.
This is significant because in addition to being able to infer the
presence of relativistic degrees of freedom, we may now also be able
to say something about the interactions of the particles which make up
that relativistic energy density.

In this paper we address the question: {\it  how much relativistic energy
density is there, and what fraction of it must consist of weakly
interacting particles?} We answer this question in general, and also
in the context of specific models.


\section{Interaction Model}
\label{sec:intmodel}

Although the results of our analysis are valid in a wider context than
the interaction model we now describe, we examine in this section a
simple physical model of non-standard neutrino interactions for
illustrative purposes.

We consider the coupling of neutrinos to each other with bosons,
through tree level scalar or pseudo-scalar couplings of the form
\begin{equation}
\label{coupl}
{\cal L}_{\nu\phi}= h_{ij}\overline{\nu}_i \nu_j \phi + 
g_{ij} \overline{\nu}_i \gamma_5 \nu_j \phi,
\end{equation}
where the boson $\phi$ is taken to be light or
massless\footnote{Couplings of neutrinos to new {\it heavy} bosons are
tighty constrained~\cite{khlopov}.}.  Such couplings arise in
Majoron-like models, viable examples of which have been discussed in
Ref.~\cite{Models}.  Recently, these models have been investigated in
the context of late-time phase transitions, whereby the neutrinos
acquire their masses via a symmetry breaking phase transition at a low
scale, which occurs late in the history of the universe
\cite{chacko,mohapatra}.  In order to be as model independent as
possible, we assume the new couplings are fixed independently of the
neutrino mass.  We also make no distinction between $g$ or $h$ type
couplings, nor between neutrinos and antineutrinos.

Existing bounds on these new couplings are extremely weak.  For
example, the solar neutrino~\cite{SolarDecay} and meson
decay~\cite{MesonDecay} limits are $|g| \lesssim 10^{-2}$.
Neutrinoless double beta decay sets a limit $g_{ee} <
10^{-4}$~\cite{doublebeta}, but does not constrain other elements of
the coupling matrix $g_{\alpha\beta}$.  Supernova constraints exclude
a narrow (and model-dependent) range of couplings around $g
\sim10^{-5}$~\cite{Supernova}.  Even couplings which are much smaller
than these limits can have significant cosmological consequences.

For a massless $\phi$ boson, scalar couplings could mediate long-range
forces with possible cosmological
consequences~\cite{LongRange,mckellar}, while pseudo-scalar couplings
mediate spin-dependent long-range forces, which have no net effect on
an unpolarized medium\footnote{For pseudo-scalar couplings, two-boson
exchange can mediate extremely weak spin-independent forces~\cite{ferrer}.}.
However, if the $\phi$ boson has even a tiny mass $H_{0} \ll m_\phi
\ll 1$ eV the interaction is short ranged and insignificant over
cosmological distance scales.

\begin{figure}[t]
\includegraphics[width=3.25in]{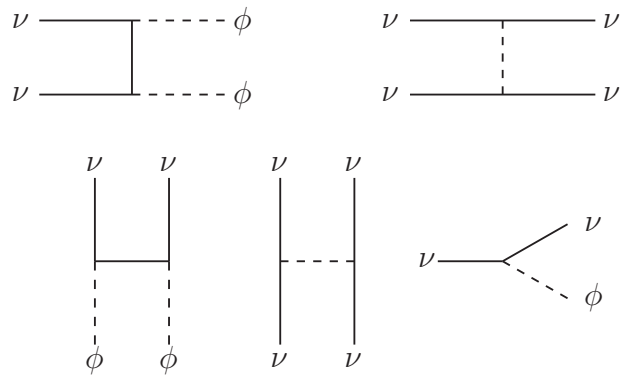}
\caption{\label{basicfig} The interactions that keep the neutrinos and
the scalar coupled.  If the scalar is heavier than $m_\nu$, the
process $\nu \leftrightarrow \nu \phi $ is replaced by $\phi
\leftrightarrow \nu \nu$.}
\label{feyn}
\end{figure}

The $\phi$ boson can be brought into thermal equilibrium through its
coupling to the neutrinos, and the $\nu - \phi$ system may stay in
thermal contact until late times.  The processes involved, shown in
Figure~\ref{feyn}, are $\nu \phi \leftrightarrow \nu \phi$, $\nu \nu
\leftrightarrow \phi \phi$, $\nu \nu \leftrightarrow \nu \nu$, and
either $\nu \leftrightarrow \nu \phi$ or $\nu \nu \leftrightarrow
\phi$, depending on whether the scalar mass, $m_\phi$, is smaller or
larger than the neutrino mass, $m_\nu$\footnote{We set all three
neutrino species to a common mass $m_\nu$, with $m_\nu \gg \sqrt{\delta
m^2_{\rm sol}}$, $\sqrt{\delta m^2_{\rm atm}}$. When this
approximation does not hold, the effects of neutrino mass are
negligible in present cosmological data.}.
For sufficiently large couplings, the $\nu$--$\phi$ system will remain
in thermal contact until the temperature drops below $m_\nu$ or
$m_\phi$.  At this point the heavier of the two particles will
annihilate or decay.

The possibility of altering the relativistic energy density through
neutrino decay has been considered in \cite{CosmoDecay}\footnote{See
also, Ref.~\cite{cuoco}, which studies the case of a scalar boson
decaying into neutrinos, thus distorting the usual thermal neutrino
distribution.
Related scenarios, in which hot dark matter
is produced by the decay of heavier particles, are examined in
Ref.~\cite{PB99}.}, while the cosmological effects of neutrino
annihilation or self-interaction were examined in
\cite{raffelt,CosmoAnnil}.  In particular, \cite{raffelt} considered
the introduction of self-interactions as a mechanism to eliminate
neutrino free-streaming, thus obtaining neutrinos which behave as cold
dark matter, despite their light mass.  However, in these scenarios
the neutrinos were taken to be heavy enough ($m_\nu \sim 10 {\rm eV}$)
to contribute all of the dark matter, which is no longer a viable
possibility.  As we will show below, the combined use of neutrino mass
limits and cosmological observations now allow much more sensitive
constraints to be placed on these types of models.

The ultimate effect of the interaction shown in Eq.~(\ref{coupl}) can
be split into two distinct cases, depending on whether the scalar is
lighter or heaver than the neutrinos.  We now discuss these cases in
turn.

\subsection{Light Scalars, $m_\phi < m_\nu$} 

When $m_\phi < m_\nu$, the neutrinos can annihilate to scalars via the
process $\nu \nu \rightarrow \phi \phi$ when the temperature drops
below $m_\nu$. Complete annihilation occurs provided that $g \agt
10^{-5}$, although smaller couplings would suffice to keep the
$\nu$--$\phi$ system in thermal equilibrium until late times via
decay/inverse decay processes.  Obtaining partial annihilation (rather
than either negligible annihilation or complete annihilation) would
require fine-tuning of $g$.

If all three neutrino species completely annihilate, they will make no
contribution to the dark matter density in the universe today, as we
would be left with a ``neutrinoless universe'' at late times.  This
eliminates cosmological constraints on neutrino
mass~\cite{neutrinoless}. However, a nonzero $\phi$ mass would make a
small contribution to the dark matter density.  Note that $m_\phi <
m_\nu$ implies the neutrinos are unstable, and may decay via $\nu_i
\rightarrow \nu_j \phi$.  For the same range of couplings, $g \agt
10^{-5}$, this could lead to neutrino decay over astronomical
distances, which is testable in future neutrino telescope
experiments~\cite{decay}.

\subsection{Light Neutrinos, $m_\phi > m_\nu$}
Alternatively, if {$m_\phi > m_\nu$, the scalar will eventually convert to
neutrinos, so that cosmological neutrino mass bounds do apply.  For
example, the decay/inverse-decay process $\phi \leftrightarrow \nu
\nu$, will keep the $\nu-\phi$ system tightly coupled throughout the
CMB era, provided $g \agt 10^{-12}$~\cite{chacko}.  Once the inverse
process becomes kinematically inaccessible at $T \sim m_\phi$, the
$\phi$ bosons will decay to neutrinos.


\section{Tightly coupled perturbation equations}

For most parameters of interest, the neutrino mean free path is much
smaller than the scales of interest during the epoch of CMB
decoupling.  In this limit, the neutrinos and bosons form a single,
tightly coupled fluid.  Perturbations in this fluid evolve
differently, compared to the usual collisionless neutrino background.
As they are coupled via gravity, neutrino perturbations influence the
evolution of photon perturbations and thus neutrino perturbations can
leave a distinctive signature in the CMB.  In this section we discuss
the relevant properties of the $\nu$--$\phi$ fluid for interesting
limits of the model considered in Section~{\ref{sec:intmodel}} and in
a more general context.

In the standard scenario, free-streaming damps the neutrino density
perturbations, and introduces a source of anisotropic stress,
e.g. see~\cite{hu,bashinsky}.  In comparison, a tightly-coupled fluid
has only density and velocity perturbations, with the shear stresses
and all higher moments in the Boltzmann hierarchy absent (at least to
linear order).  Defining the density and velocity perturbations as
$\delta = \delta \rho/\rho$ and $\theta = ikv$ respectively, the
equations describing the evolution of the tightly-coupled
$\nu$--$\phi$ fluid are
\begin{eqnarray}
\label{eqn:perturbations}
\dot{\delta} &=& -(1+ \omega) \left( \theta + \frac{\dot{h}}{2} \right)
-3 \frac{\dot{a}}{a}\left( c_s^2 - \omega \right) \delta, \\
\dot{\theta} &=& -\frac{\dot{a}}{a}\left( 1 - 3 \omega \right) \theta
- \frac{\dot{\omega}}{1+\omega} \theta + \frac{c_s^2}{1+\omega}k^2\delta,
\end{eqnarray}
in the synchronous gauge, where an overdot is a derivative with
respect to conformal time.  Here, $c_s^2 = \delta P/ \delta \rho$ is
the adiabatic sound speed and $\omega = P/\rho$ is the equation of
state with $\rho = \rho_\nu + \rho_\phi$ and $P = P_\nu + P_\phi$.  In
the limit where both $\nu$ and $\phi$ are relativistic $\omega= c_s^2=
1/3$.  However, for non-zero masses, $\omega$ and $c_s^2$ temporarily
decrease from $1/3$ and deviate from each other, during the period
when $\nu$ or $\phi$ starts to become non-relativistic and
annihilate/decay.  We may define an effective number of standard model
neutrinos $\Nnuint$ contributed by the $\nu$--$\phi$ fluid with the
relation
\begin{align}
\rho=\Nnuint\frac{7}{8}\frac{\pi^2}{15}\left(T_{\nu}^{\mbox{\tiny SM}}\right)^4, \,
\label{eqn:energydensity}
\end{align}
where $\TnuSM=(4/11)^{1/3}T_{\gamma}$ is the canonical cosmic neutrino
background temperature in the standard model.  The value of $\Nnuint$
after the heavier species annihilates into the lighter will be greater
than the value before annihilation.  Details of the calculation of
$w$, $c_s^2$, and $\Nnuint$ can be found in Appendix~\ref{App:A}.

The differences from standard cosmology in these models will therefore
typically be a combination of (i) elimination of neutrino
free-streaming, (ii) nonstandard equation of state and sound speed
evolution, and (iii) additional (and perhaps an evolving amount of)
relativistic energy density.  We now consider two phenomenological
models (limiting cases of the interaction model of
Section~\ref{sec:intmodel}) that exhibit these differences.

\subsection{Model A:  
Freestreaming vs. Interacting (Massless Particles) }\label{constantN}

In this section we consider a scenario in which some fraction of the
neutrinos act as a tightly-coupled fluid, but where $\Neff$ is
constant with time.  
We parameterize this class of models by $\NnuSM$ and $\Nnuint$, the
number of standard model (free-streaming) and interacting degrees of
freedom respectively, where 
\begin{equation}
\Neff=\NnuSM+\Nnuint.
\end{equation}  
A scenario in which $\Neff$ is constant in time corresponds, for example, to
the limit $m_\nu \rightarrow 0$ and $m_\phi \rightarrow 0$.  However,
even for finite masses, this limit is a good approximation as long as
$\Neff$ does not evolve significantly during the times of interest, as
would be the case if any annihilation/decay takes place well before or
well after the CMB decoupling epoch.  In particular, this is a
reasonable approximation for the $m_\phi > m_\nu$ models considered in
\cite{chacko}, where only modest evolution of $\Neff$ occurs.

Note that this description is much more general than any particular limit
of a neutrino-scalar interaction model.  It encompasses any scenario
in which some fraction of the energy density is in free-streaming
relativistic particles, and another fraction is in a tightly-coupled
relativistic fluid, as long as $\Neff$ does not evolve significantly.

\begin{figure}[t]
\includegraphics[width=3.35in]{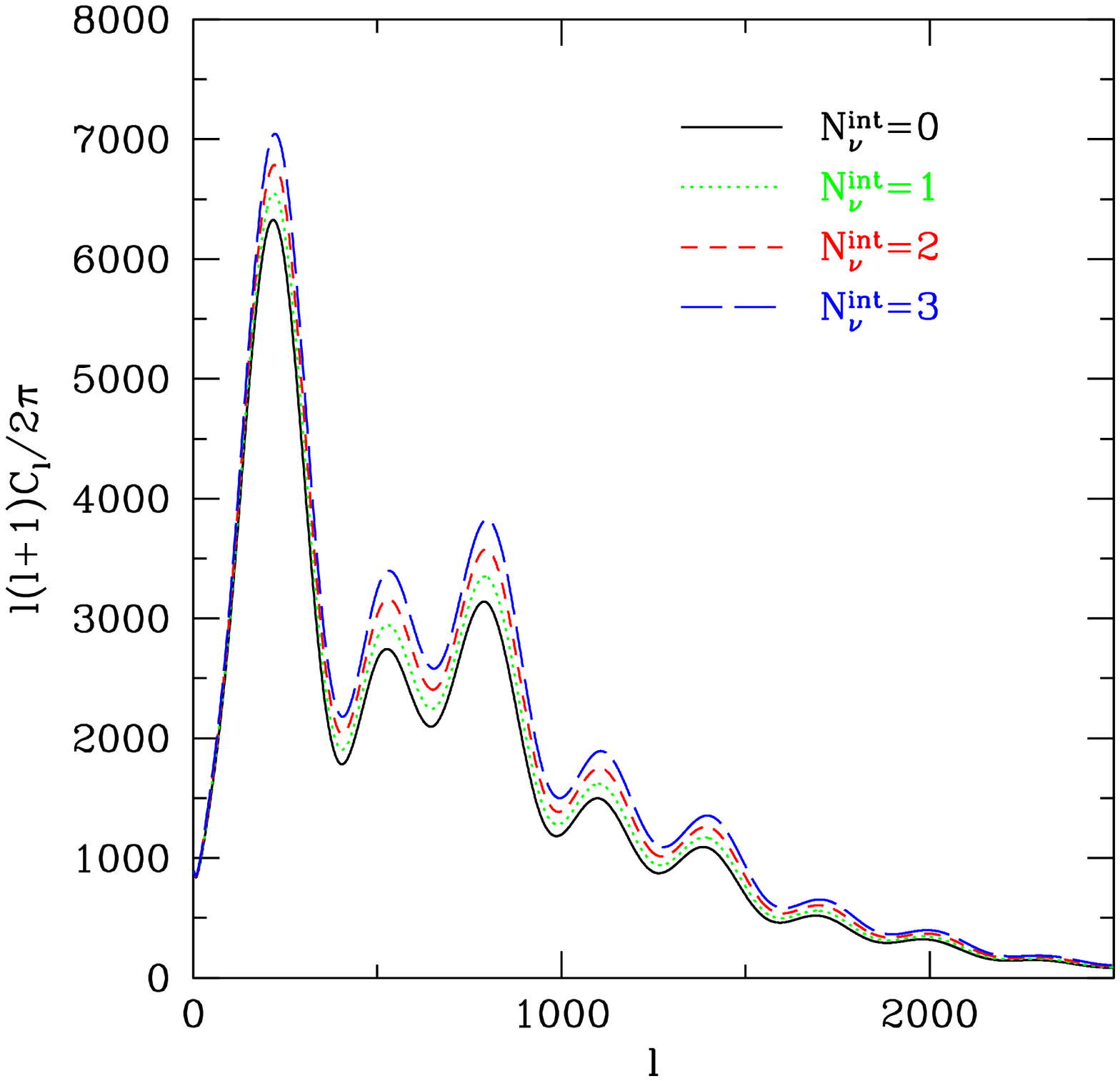}
\includegraphics[width=3.35in]{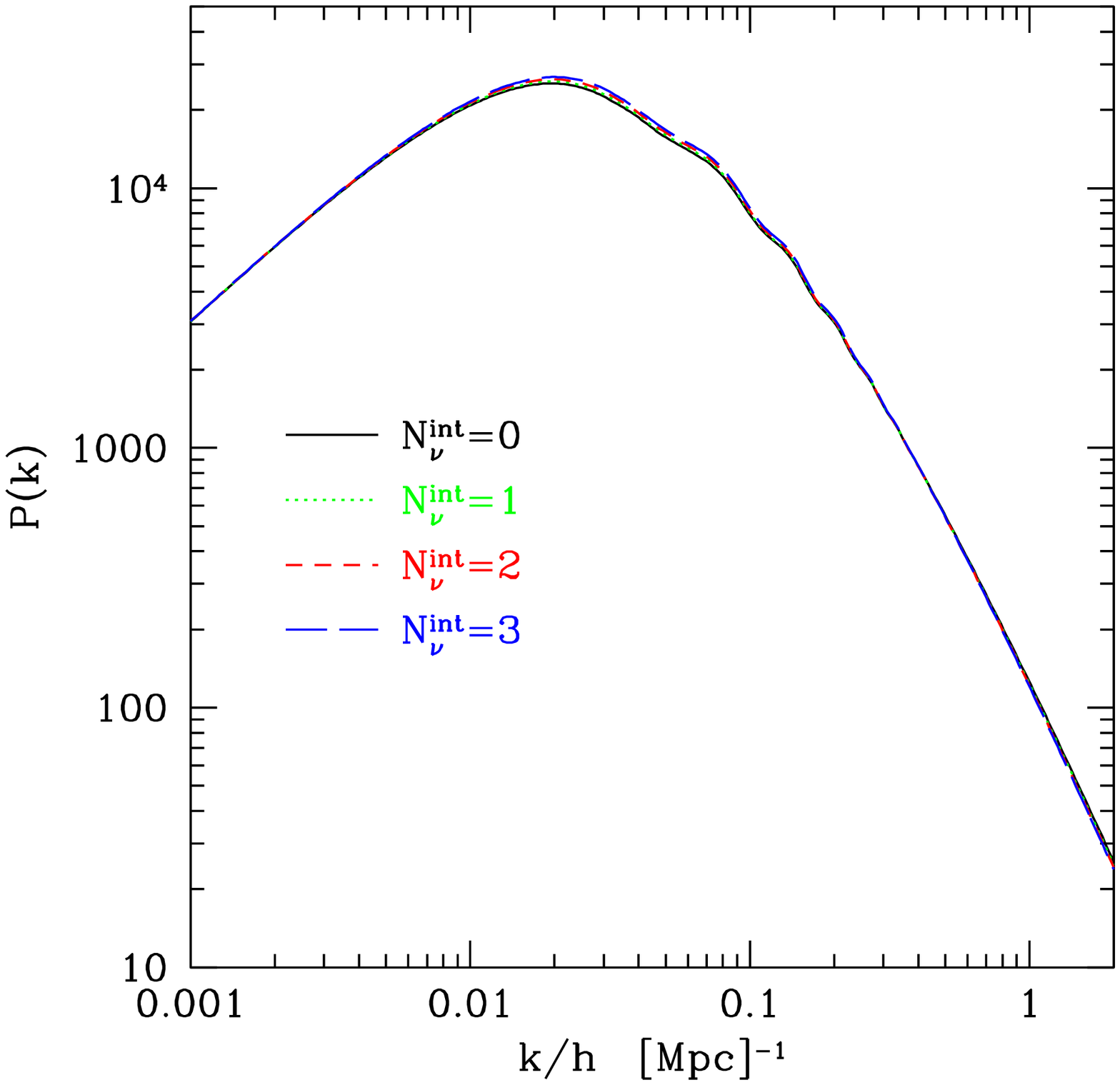}
\caption{\label{fig:fint_cl} CMB and matter power spectra as a
function of the fraction of interacting neutrinos, with
$\Neff \equiv \NnuSM+\Nnuint=3$.
The power spectra are normalized (to an arbitrary value) at large scale.}
\end{figure}


\begin{figure}[t!]
\includegraphics[width=3.35in]{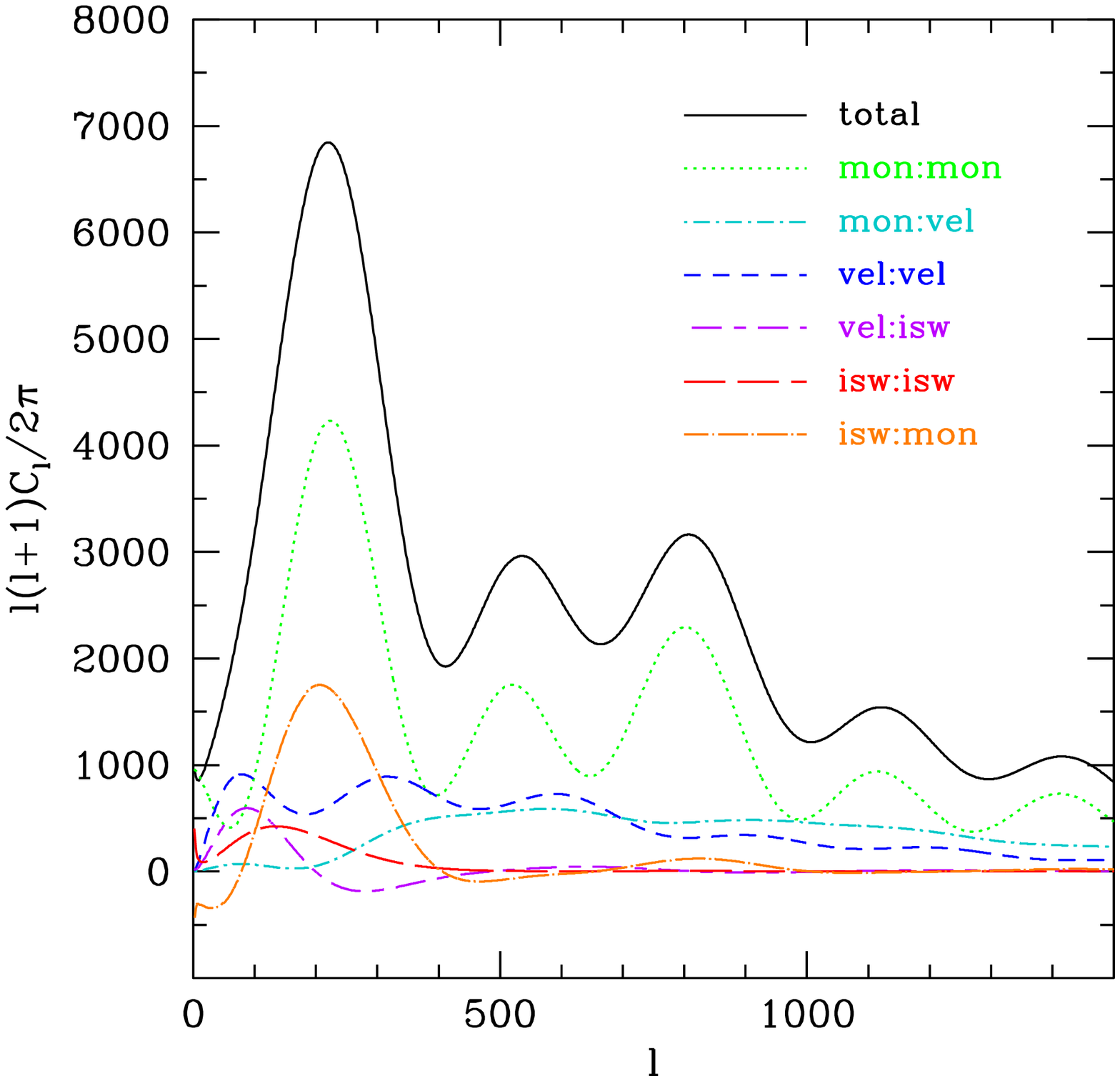}
\includegraphics[width=3.35in]{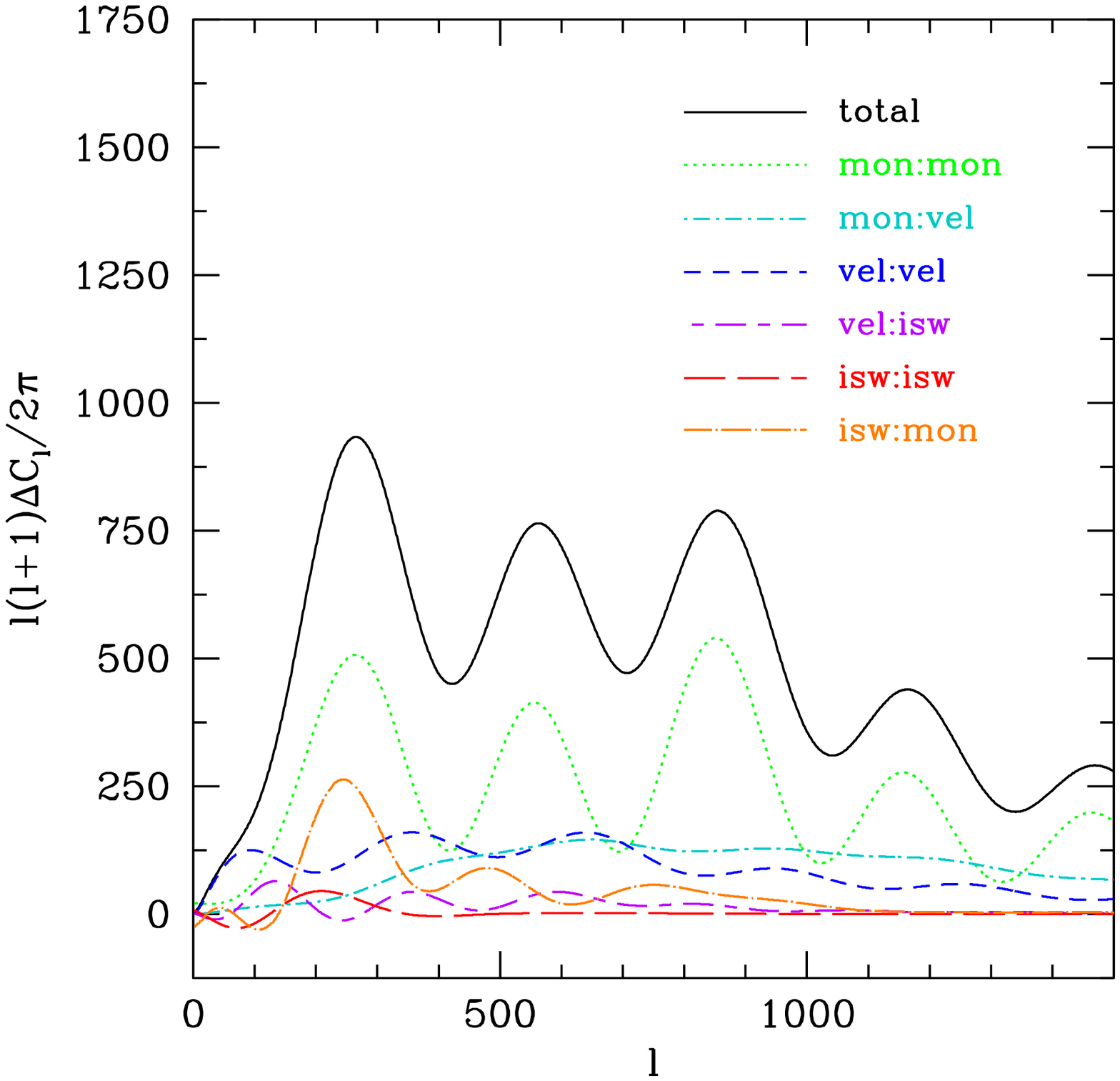}
\caption{\label{fig:source_terms} Upper: The CMB
power spectra due to the monopole, velocity, and ISW terms of the source function for the standard $\Nnuint=0$ case (For a pedagogical description of these terms see, e.g., Refs.~\cite{Seljak:1996is,Dodelson:2003ft}).  Lower:  The contribution of each source term to $\Delta C_l$, the difference between a model with $\Nnuint=3$ and $\Nnuint=0$ with $\Neff=3$ held fixed.  The ISW-monopole cross term (nonzero due to the differing evolution of the gravitational potentials) contributes significantly to the difference in the first peak despite the identical background evolution of the two models.}
\end{figure}

\begin{figure}[t]
\includegraphics[width=3.35in]{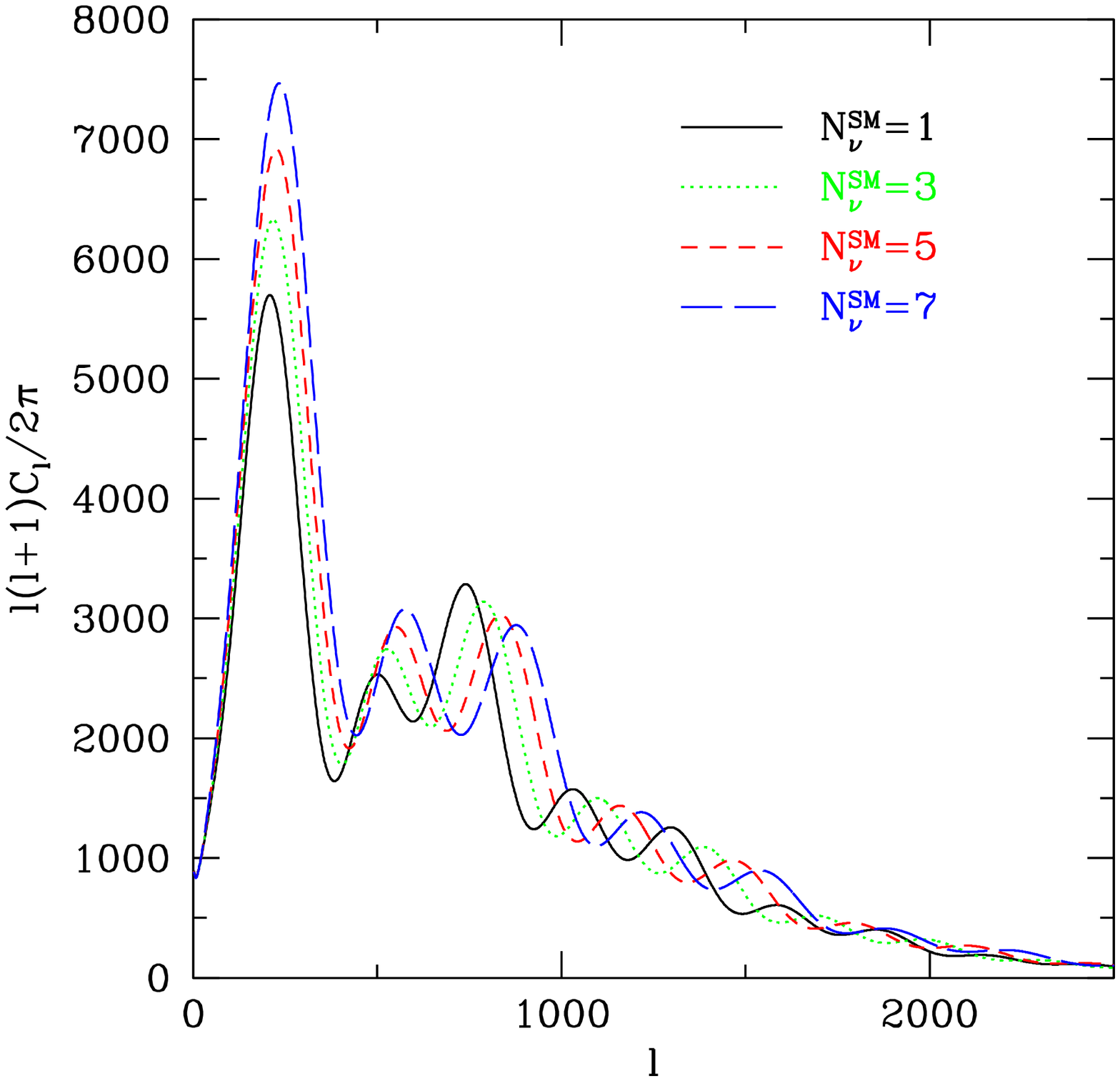}
\includegraphics[width=3.35in]{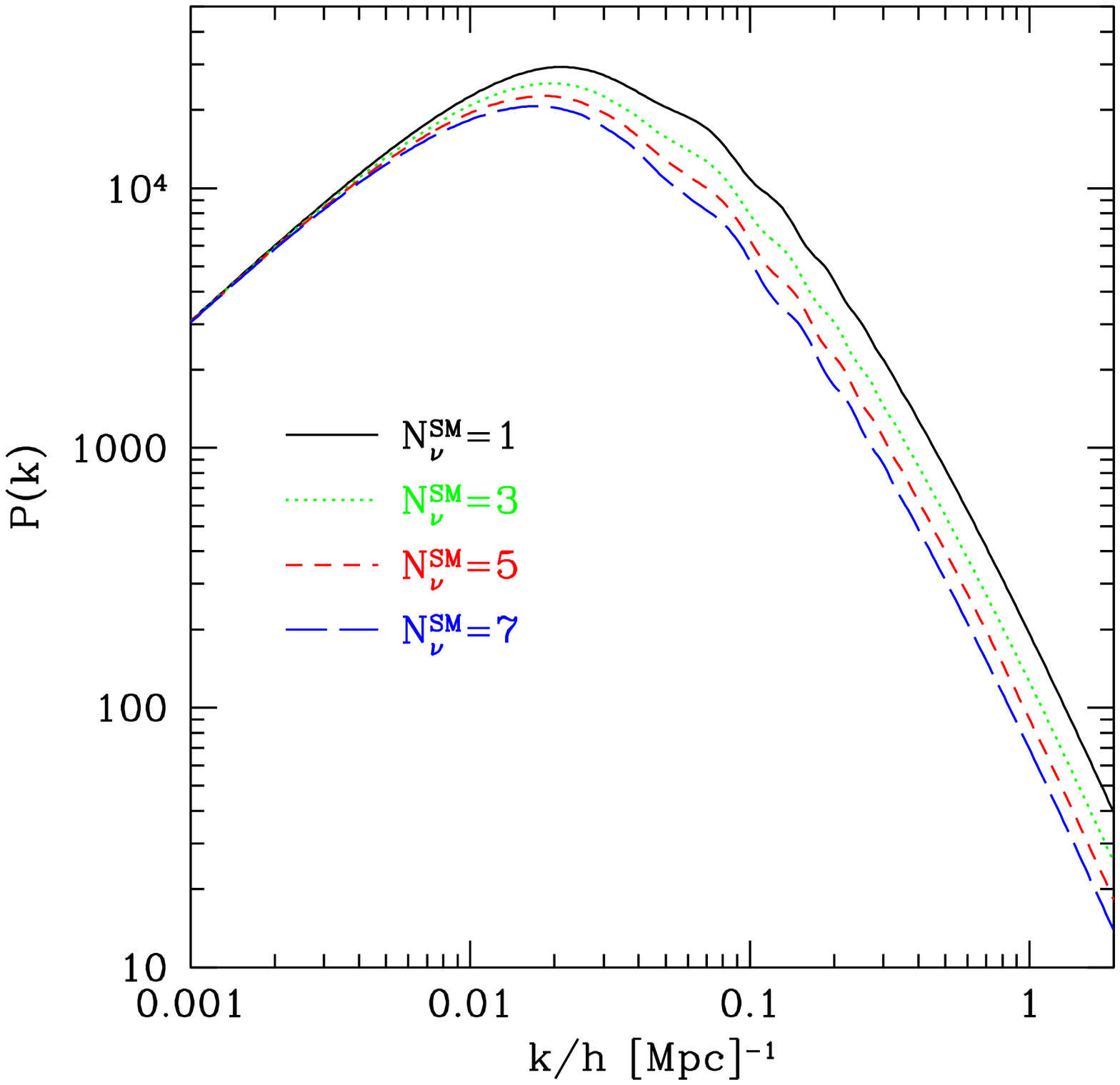}
\caption{\label{fig:nnu_cl} The CMB and matter power spectra as a
function of the number of standard model neutrinos, with  $\Nnuint=0$.
The power spectra are normalized (to an arbitrary value) at large scale.}
\end{figure}

Fig.~\ref{fig:fint_cl} shows effects on the CMB angular power spectrum
and the matter power spectrum of varying the number of interacting
neutrinos, while $\Neff$ and all other cosmological parameters are
held fixed.  This allows us to explore the effects of suppressed
neutrino free streaming alone.  In the standard case ($\Nnuint=0$) 
when a perturbation of a given scale enters the horizon, power
is transfered from the neutrino density modes to higher moments of 
the neutrino distribution as the neutrinos free-stream out of 
gravitational potentials.  This effect does not occur for 
interacting neutrinos which instead contribute to the gravitational 
potential and thus enhance the monopole perturbation of the photon distribution and increase the amplitude of the CMB temperature power spectrum for a fixed amplitude of primordial perturbations.  Indeed, we see that beyond the 
first peak the effect of increasing $\Nnuint$ in the CMB can
be roughly approximated as a constant offset in the amplitude of the
spectrum. A small shift in the location of the peaks is also a
feature~\cite{bashinsky}.  In the upper panel of Fig.~\ref{fig:source_terms} we show the contributions to the CMB power spectrum arising from the monopole terms (the physical temperature perturbation at the last-scattering surface corrected for its gravitational redshift), the velocity terms (perturbations due to the Doppler shift), and the ISW terms (perturbations due to the evolving gravitational potentials) of the source function  (see, for example, Refs.~\cite{Seljak:1996is,Dodelson:2003ft} for a pedagogical review of the source function).  The lower panel of Fig.~\ref{fig:source_terms} shows the difference in CMB power spectra between models with  $\Nnuint=3$ and $\Nnuint=0$ keeping $\Neff=3$ fixed (the extreme models of Fig.~\ref{fig:fint_cl}). 
While the change in the monopole dominates the total difference in 
the power spectrum at all $l$, the ISW-monopole cross term 
contributes significantly to the difference near the first peak.
Note that this ISW-monopole contribution is nonzero solely 
because gravitational potentials evolve differently in a model
with and without free-streaming neutrinos --- the background
evolution is identical in the two cases.

As we are in the limit of massless neutrinos, the effect on the matter
power spectrum of suppressing free streaming is very minor.
Free-streaming has a significant effect on the matter power spectrum
when $m_\nu$ is finite, such that neutrinos contribute some fraction
of the dark matter density today.  In that case, the free-streaming of
the neutrino hot dark matter component damps the growth of structure
while the neutrinos are still relativistic.  By comparison, the effect of
modifications to the neutrino perturbations of massless neutrinos is
very small.

For comparison, we show in Fig.~\ref{fig:nnu_cl} the effects on the CMB
angular power spectrum and the matter power spectrum of varying the
total relativistic energy density, $\NnuSM$, while $\Nnuint$ and all
other cosmological parameters are held fixed.  We see that increasing
$\Neff$ enhances the first peak.  This is the result of a larger ISW
effect, due to the delay in matter-radiation equivalence.  There is
also a large shift in the positions of the subsequent peaks, which
occurs due to the change in the conformal time of last scattering.  
See Ref.~\cite{Pierpaoli} for a discussion of these effects.
In Fig.~\ref{fig:nnu_cl}, we also see that the enhanced radiation density
suppresses the matter power spectrum.  Again, this is because
matter-radiation equivalence, and hence the growth of structure, is
delayed.

The extent to which the various effects discussed above can
be compensated with a change in other cosmological parameters will be
discussed in section~\ref{results}.



\subsection{ Model B: Neutrino Annihilation to Scalars}\label{annihilation}

We now turn to the slightly more complicated scenario with $m_\phi <
m_\nu$, where the interacting neutrinos can annihilate.  We again assume the 
limit $m_\phi \rightarrow 0$, but allow a nonzero $m_\nu$.  
We specialize to the case of three neutrino species,
and will consider the possibility that either one, two, or all three
neutrino species interact strongly with the scalar $\phi$, which we
shall denote by models B1, B2, and B3 respectively.  For couplings
constants $g \agt 10^{-5}$, the neutrino species which are coupled to
the scalar will annihilate when $T_\nu \sim m_\nu$.  If all three neutrino
species annihilate (model B3) this leaves a ``neutrinoless universe''.
Cosmological neutrino mass bounds are altered in these scenarios,
because the neutrino species which annihilate will not make a
contribution to the dark mater density today (i.e., they will not
contribute to $\Omega_\nu$)~\cite{neutrinoless}.

\begin{figure}[t]
\includegraphics[width=3.10in]{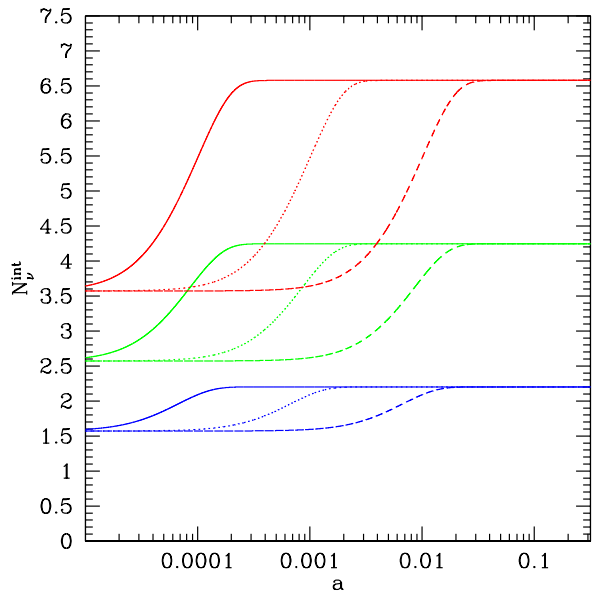}
\caption{\label{Neff} The evolution of the effective number of
interacting neutrinos $\Nnuint$ with $m_{\nu}=0.1$~eV (dashed curves),
$m_{\nu}=1$~eV (dotted curves) and $m_{\nu}=10$~eV (solid curves) for
3 interacting neutrinos (top/red), 2 interacting neutrinos
(middle/green), and 1 interacting neutrino (bottom/blue).  Notice that
the $\Nnuint$ initially includes an extra $4/7$ to account for the new
scalar degree of freedom and that the effective number of interacting
neutrinos after annihilation is greater than prior to annihilation.}
\label{fig:Neff}
\end{figure}
\begin{figure}[h]
\includegraphics[width=3.10in]{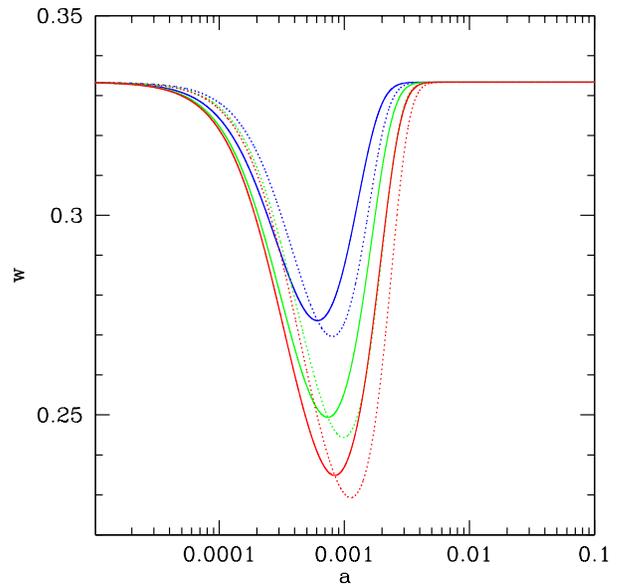}
\caption{\label{w_cs2} The evolution of $w$, the equation of state
(solid curves) and $c_{s}^2$, the sound speed (dotted curves) as a
function of the scale factor, $a$, for $m_\nu = 1$~eV for 3
interacting neutrinos (bottom/red), 2 interacting neutrinos
(middle/green), and 1 interacting neutrino (top/blue).  Both deviate
from the value of $1/3$ for a relativistic fluid, during the epoch of
annihilation. }
\label{fig:w}
\end{figure}

A distinctive feature of this scenario is that the neutrino
annihilation will heat the scalars, causing $\Neff$ to evolve as the
annihilation proceeds.  For simplicity, we shall assume the scalar
boson is brought into thermal equilibrium before the neutrinos
thermally decouple from the electrons and positrons (at $T \sim 1$ MeV).  The scalar will
then initially contribute an amount $\delta N_\nu = 4/7$ to the
relativistic energy density, so that $\Neff \simeq 3.57$
\footnote{If the scalar were not populated until sometime after the
neutrinos thermally decouple from the $e^+e^-$ plasma, $\Neff$ would
not be altered (as energy density would simply be shifted from one
relativistic species to another.)}.  This value will increase as the
annihilation proceeds.  For example, if all three neutrinos annihilate,
the final relativistic energy density is equivalent to $N_\nu \simeq
6.6$~\cite{neutrinoless} (see also Appendix~\ref{App:A}).  For
realistic neutrino masses, $m_\nu \simeq 0-2$ eV, this
annihilation occurs close to the time at which the CMB photons last
scatter, so that the evolution of $\Neff$ takes place during the CMB
decoupling era.  The evolution of $\Neff$ is shown in
Fig.~\ref{fig:Neff}.  The evolution of the equation of state, $w$, and
sound speed, $c_{s}^2$, during this annihilation epoch is shown in
Figure.~\ref{fig:w}.

The present laboratory limit on neutrino mass is $m_\nu < 2.2 {\rm
~eV}$, set by tritium beta decay experiments~\cite{tritium}.  Given
the tiny mass squared differences measured by solar and atmospheric
neutrino oscillation experiments ($\delta m^2_{\rm sol} \simeq 7
\times 10^{-5} {\rm ~eV}^2$ and $\delta m^2_{\rm atm} \simeq 2 \times
10^{-3} {\rm ~eV}^2$~\cite{oscillation}), the tritium bound applies to
all three neutrino mass eigenstates.  We shall assume that the three
neutrino eigenstates have degenerate masses, which is a good
approximation for $m_\nu \agt 0.1 {\rm ~eV}$.  (We use $m_\nu$ to
denote the value of a single neutrino mass throughout, so that the
quantity $\Sigma m_\nu = 3 m_\nu$.)  In the analysis in
section~\ref{results}, we consider neutrino masses in the allowed
range $m_\nu = 0-2.2 {\rm ~eV}$.

For simplicity, we also assume that the scalar has sufficiently strong
self-interactions that it continues to behave as a tightly-coupled
fluid once the neutrino annihilation is complete.  For early neutrino
annihilation (large masses) this will result in the largest deviations
of the CMB spectra with respect to the standard scenario.  For late
annihilation (small masses) the late-time behavior of the scalar is
irrelevant.

\begin{figure}[t]
\includegraphics[width=3.2in]{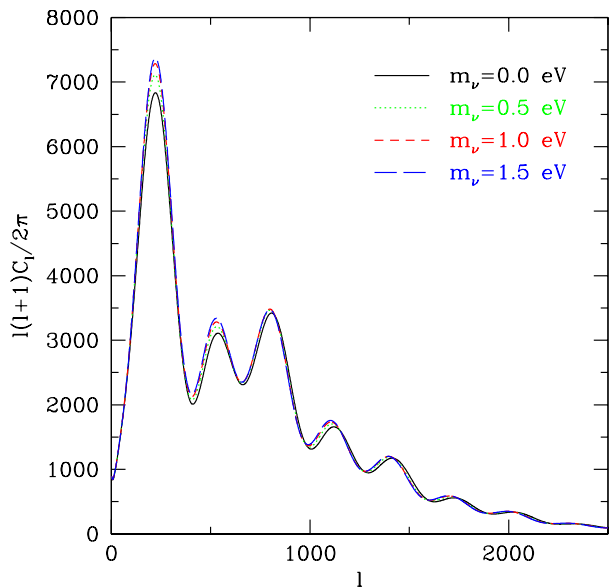}
\includegraphics[width=3.2in]{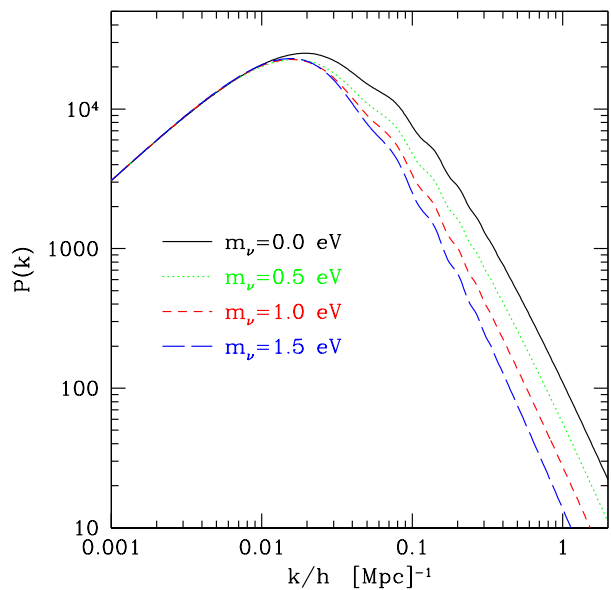}
\caption{\label{fig:na1_cl} The CMB and matter power spectra, for
varying values of $m_\nu$, for model B1 (one interacting neutrino and
two standard neutrinos).  The power spectra are normalized (to an
arbitrary value) at large scale.}
\end{figure}
\begin{figure}[t]
\includegraphics[width=3.2in]{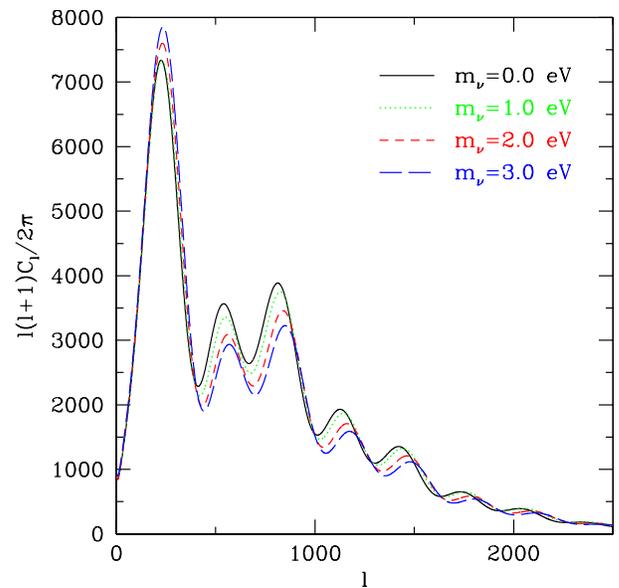}
\includegraphics[width=3.2in]{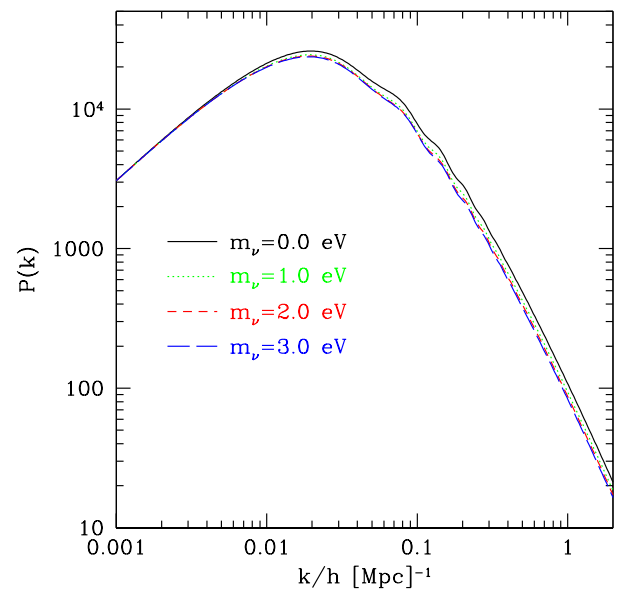}
\caption{\label{fig:nb3_cl} The CMB and matter power spectra, for
varying values of $m_\nu$, for model B3 (three
interacting/annihilating neutrinos).  The power spectra are
normalized (to an arbitrary value) at large scale.}
\end{figure}

Fig.~\ref{fig:na1_cl} shows effects on the CMB angular power spectrum
and matter power spectrum of varying $m_\nu$ in the model with one
interacting neutrino (B1), while Fig.~\ref{fig:nb3_cl} shows the same
effects for the model with three interacting neutrinos (B3).
Here we see a combination of the effects of free-streaming suppression
(compare Fig~\ref{fig:fint_cl}) and larger $\Neff$ (compare
Fig.~\ref{fig:nnu_cl}).  In the CMB spectra, we see the enhanced
overall amplitude which is characteristic of the tightly coupled
neutrinos, together with an enhanced first peak and shifted subsequent
peaks that result from increased relativistic energy density.
As expected, the effects on the CMB spectra are more pronounced in model
B3, in which all three neutrinos species annihilate.  Moreover, the
effect increases as $m_\nu$ increases, because larger mass corresponds
to earlier annihilation, so that the extra relativistic energy density
is present earlier.

Conversely, the effects on the matter power spectra are more
pronounced in model B1 in which only one neutrino species annihilates.
This is expected, because the remaining neutrinos contribute to the
dark matter density today.  This remaining neutrino hot dark matter
component causes the usual suppression of the power spectrum that is
used to constrain neutrino mass.
However, in the ``neutrinoless'' model (B3), this suppression of the
power spectrum is absent, because no neutrino dark matter component
remains.  A smaller power spectrum suppression does remain, as shown
in Fig.~\ref{fig:nb3_cl}.  This small effect is a result of enhanced
$\Neff$, which slightly delays matter-radiation equivalence and hence
delays the growth of structure.  Again, this effect is more
significant for larger $m_\nu$, because the extra radiation is present
earlier.
We thus see that the ``neutrinoless universe'' model (B3) can accommodate 
large neutrino mass, while having little effect on the matter power
spectrum~\cite{neutrinoless}.

\section{Data}

We computed the CMB and large scale structure power spectra, and performed
the parameter estimation with a modified version of the publicly available 
Markov Chain Monte Carlo package, COSMOMC \cite{COSMOMC,CAMB}.

Constraints on the models were evaluated using CMB data from the
WMAP~\cite{WMAP}, ACBAR~\cite{ACBAR} and CBI~\cite{CBI} experiments,
together with the galaxy power spectra measured by the 2dF Galaxy
Redshift Survey~\cite{2dF} and the Sloan Digital Sky Survey
(SDSS)~\cite{SDSS}.  In addition, we used the measurement of the
Hubble parameter made by the Hubble Space Telescope (HST) Key
Project~\cite{HST}. This set of data constitutes our basic comparison
set (hereafter \emph{Cosmo}).  For model B3, we imposed a cut-off of
2.2 eV for the neutrino mass, consistent with the tritium beta decay
limit~\cite{tritium}.

In addition, we will investigate how the introduction of the
Lyman-$\alpha$ constraints \cite{lyman} and the CMB polarization as
measured by CBI \cite{CBIpol} constrain these models.  These two
variants will be named \emph{Cosmo}Ly$\alpha$ and \emph{Cosmo}CBIpol
respectively.  We implement the Lyman-$\alpha$ constraints derived in
Ref.~\cite{VHS04} and we implement them in a similar way as in
Ref.~\cite{lyman}, with minor modifications that were suggested by the
authors.
 
Note that SDSS and 2dF data impose a constraint only on the shape of the
power spectrum, $P(k)$, and not on the normalization, as we make no
assumptions about the bias.  Lyman-$\alpha$ data, however, imposes constraints
on both normalization and shape at small scales.


\section{Results}\label{results}

We now discuss how the different models fit the data.  We first note
that for all of the models, parameters sets can be found which provide
a good global fit to the data, and so none of the models can be ruled
out.  For example, in Table~\ref{tab:chi2} we report the $\chi^2$
values of the best fit points in our Markov chains for the case of the
\emph{Cosmo} data set.  The table shows that the best fit parameter
set for all the models considered indeed provides a good fit to the
data.  We plot spectra for the best fit models in
Fig.~\ref{fig:bestfit}.  Apart from the high value of $\NnuSM$ in
model A (see the discussion at the end of the the next section), the
best-fit parameters are within commonly adopted parameter ranges in
the standard $\Lambda$CDM model.

\begin{figure*}[t]
\includegraphics[width=3.15in]{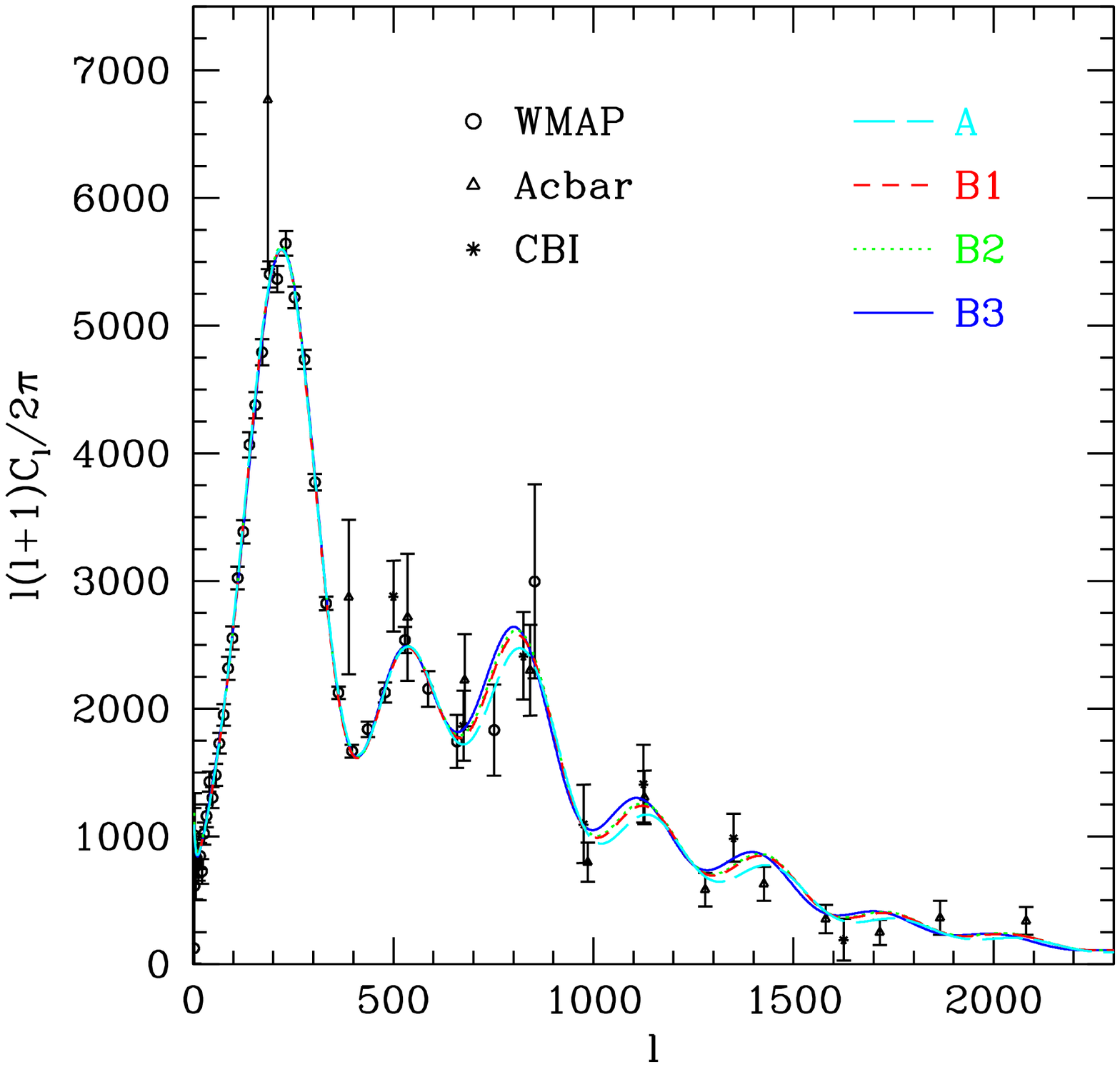}
\hspace{0.5cm}
\includegraphics[width=3.15in]{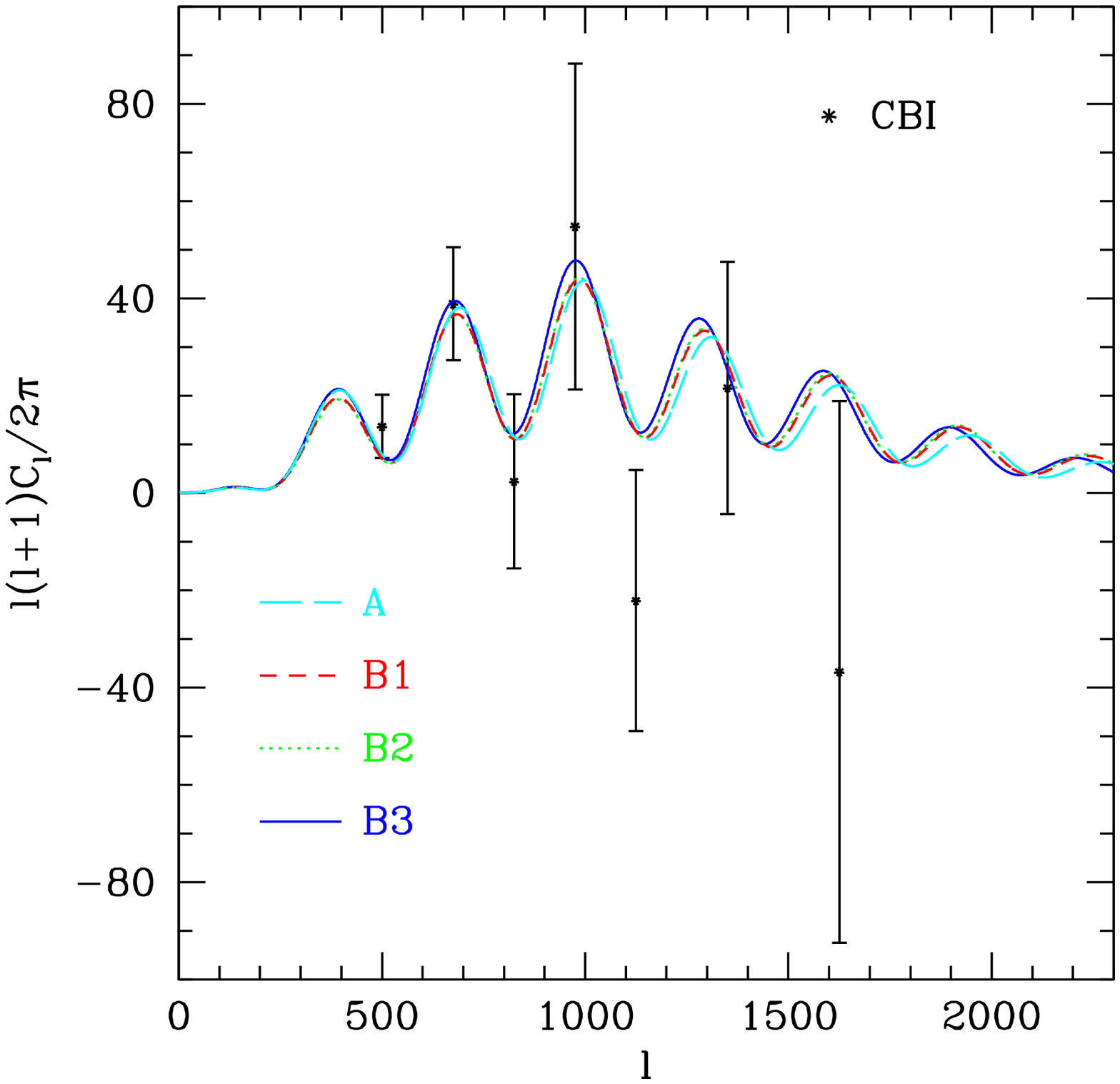}
\includegraphics[width=3.15in]{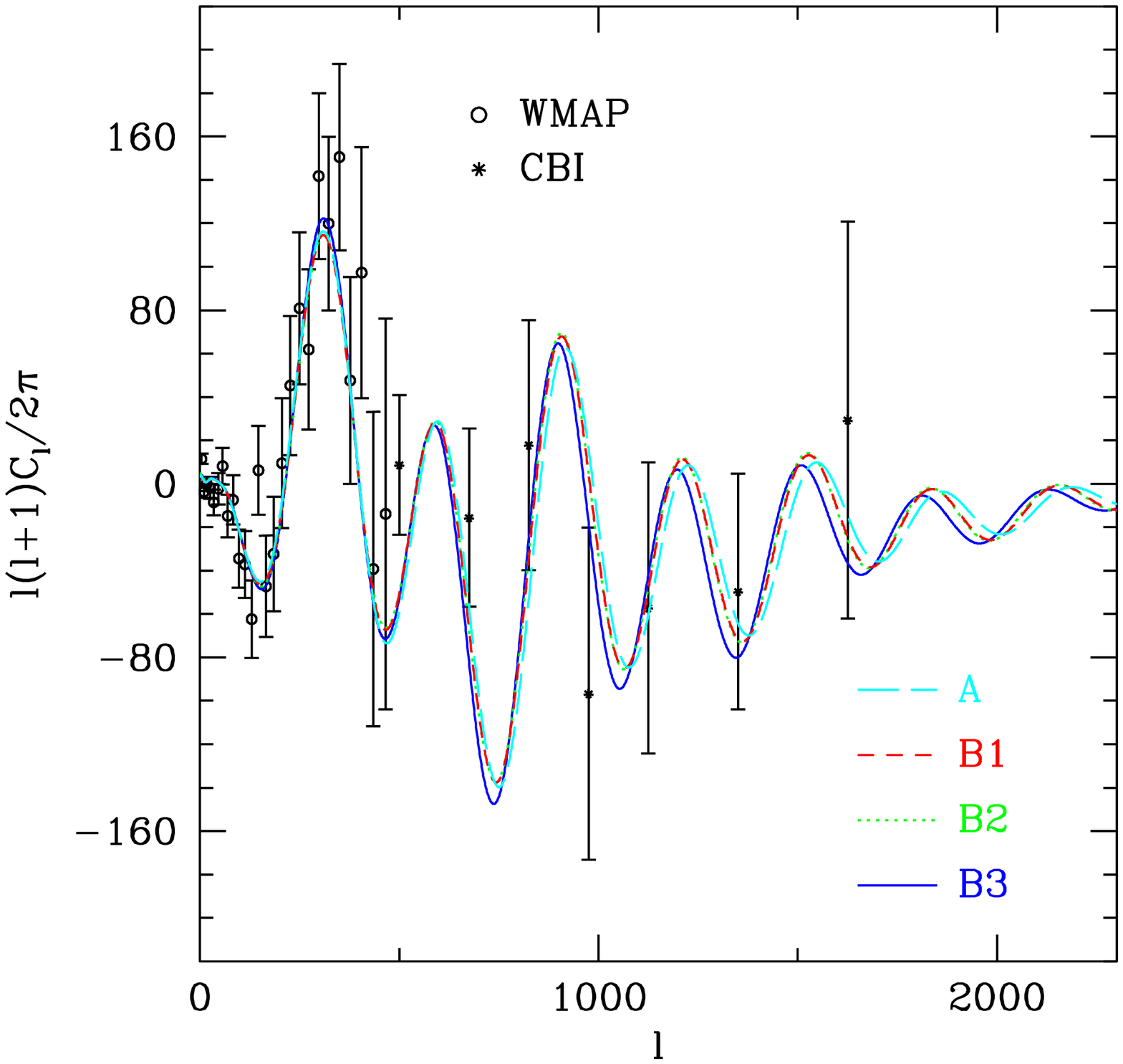}
\hspace{0.5cm}
\includegraphics[width=3.15in]{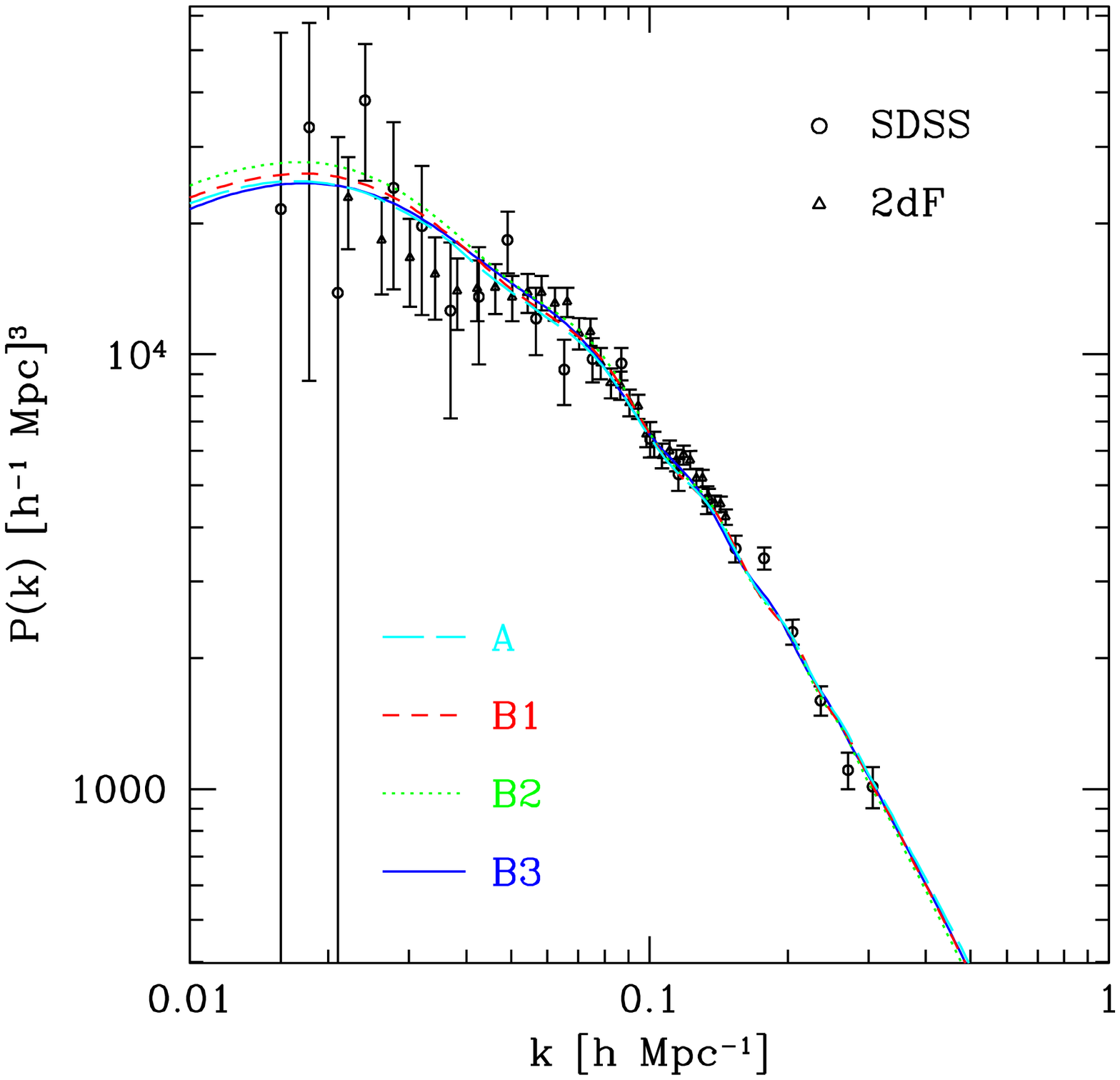}
\caption{\label{fig:bestfit} 
The CMB temperature (TT, upper left), polarization (EE, upper right) and
cross (TE, lower left) power spectra, and the matter power spectrum (lower
right), for the best-fit models.  In each case models can be found that
fit all data well.}
\end{figure*}
\begin{table}[h]
   \centering
   \begin{tabular}{c|c|c|c|c}
 \hline\hline
      Model & $\chi^2$ &  DOF & Reduced $\chi^2$ & $({\rm DOF}/2)^{-1/2}$ \\ 
\hline\hline
      A  & 1515.00 & 1478 & 1.025  & 0.037 \\
      B1 & 1520.17 & 1479 & 1.028  & 0.037 \\
      B2 & 1523.99 & 1479 & 1.030  & 0.037 \\
      B3 & 1526.77 & 1479 & 1.032  & 0.037 \\
\hline\hline
   \end{tabular}
   \caption{The $\chi^2$ values and degrees of freedom (DOF), using
   the \emph{Cosmo} data set.}
   \label{tab:chi2}
\end{table}

On the other hand, while reasonable parameters can be found that yield
a good global fit to the data for each of model A, B1, B2, and B3, we
will show that models with fewer interacting neutrinos are preferred
in a Bayesian sense after marginalizing over all other cosmological
parameters.  The interpretation of these results is discussed further
below.
We shall now discuss each model individually, focusing on the 
constraints on the neutrino properties.

\subsection{ Model A: $\NnuSM$ vs. $\Nnuint$}

Fig.~\ref{ninnimarg} displays the curves for the marginalized
likelihood of the parameters $\NnuSM$ and $\Nnuint$.  Notice that the CMB
alone allows a larger number of interacting neutrinos, while preferring
a relatively low number of standard neutrinos.  The addition of the matter
power spectra yields the effect of reducing the maximum value of
$\Nnuint$, and also shifts the peak of the likelihood for $\NnuSM$.
Referring to Fig.~\ref{fig:fint_cl}, we notice that keeping the total
number of neutrinos fixed while increasing the number of interacting
neutrinos has a large impact on the CMB power spectrum, and a very minor one
on the matter power spectrum (the two models have equal
total energy density, but different neutrino fluctuation evolution).
However, the introduction of the matter power spectrum constraint in the
likelihood analysis imposes limits on the epoch of equivalence, and on
the total spectral index.  This, in turn, leaves less freedom for
accommodating non--standard neutrino fluctuation evolution which
affects the CMB power spectrum.

\begin{figure}[t]
\includegraphics[width=3.35in]{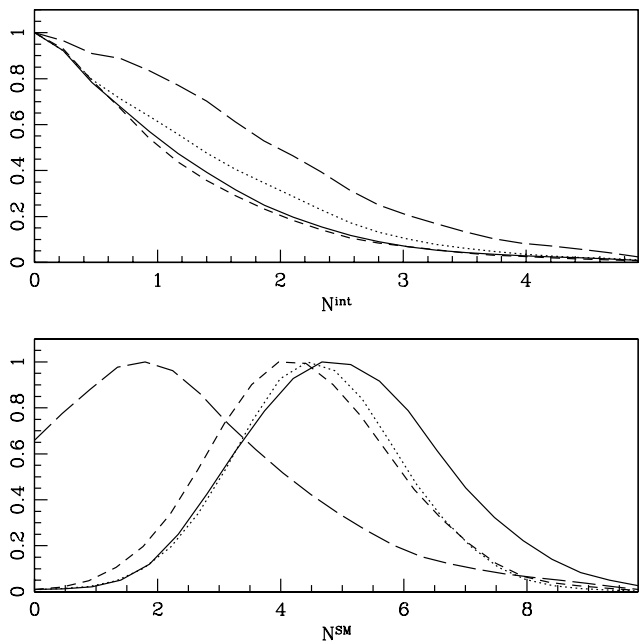}
\caption{The marginalized likelihood curves for $\Nnuint$ and $\NnuSM$
in Model A.  The solid curves corresponds to the \emph{Cosmo} data set
(CMB+2dF+SdDS+HST) and the long-dashed curve to the CMB data alone.
The dotted and short-dashed curves correspond to the addition of the
Lyman-$\alpha$ and CBI polarization data respectively.}
\label{ninnimarg} 
\end{figure}

As anticipated in~\cite{Pierpaoli}, the addition of the Lyman--$\alpha$
constraint improves the limits on the standard model neutrinos.
This is because by probing the power of the fluctuations on very small
scales, the Lyman--$\alpha$ data places a strong constraint on the
spectral index $n$ \cite{lyman}, which is degenerate with $\NnuSM$ and
the reionization $\tau$ in the radiation power spectrum.

It is interesting to note that adding the CBI polarization data also
improves the constraints on $\NnuSM$.  This is because an increased
number of neutrinos significantly shifts the peaks of the $C_l$, due
to the delay in matter-radiation equivalence.  The CBI polarization
data have been shown to be able to determine the phase of the
oscillations with high precision \cite{CBIpol}, despite the size of
the errorbars.  This is one example of how such information can be
used in constraining parameters.

Table \ref{tab:ninni} summarizes the best fit values and the 95 \%
confidence levels for the marginalized likelihoods for $\NnuSM$ and
$\Nnuint$, while Fig.~\ref{ninni} shows the degeneracy between
$\NnuSM$ and $\Nnuint$ in the case of the \emph{Cosmo} dataset.  A
higher value for $\Nnuint$ may be compensated by a lower value for
$\NnuSM$.  However, once the matter power spectrum is used in the
analysis (and thus the redshift of equivalence is constrained) the
degeneracy is mild.

\begin{table}[t]
   \centering
   \begin{tabular}{c||c|c||c|c} 
\hline\hline
     & $\NnuSM$ &  $\NnuSM$ &  $\Nnuint$ &  $\Nnuint$ \\
Data set   & best fit &  upper limit &  best fit &  upper limit \\
\hline\hline
      \emph{Cosmo}  &  5.0  &   7.8 &  0.015 &  2.9 \\
  \emph{Cosmo}CBIpol  &  4.3  &  6.9   & 0.1  & 2.9  \\
   \emph{Cosmo}Ly$\alpha$  & 4.9    &  6.8  &  0.02 &  3.0 \\
\hline\hline
       \end{tabular}
\caption{The best fit values and 95 \% C.L. allowed ranges for
$\NnuSM$ and $\Nnuint$ in Model A. The confidence limits are obtained
from marginalized curves.}
   \label{tab:ninni}
\end{table}

\begin{figure}[h]
\includegraphics[width=3.0in]{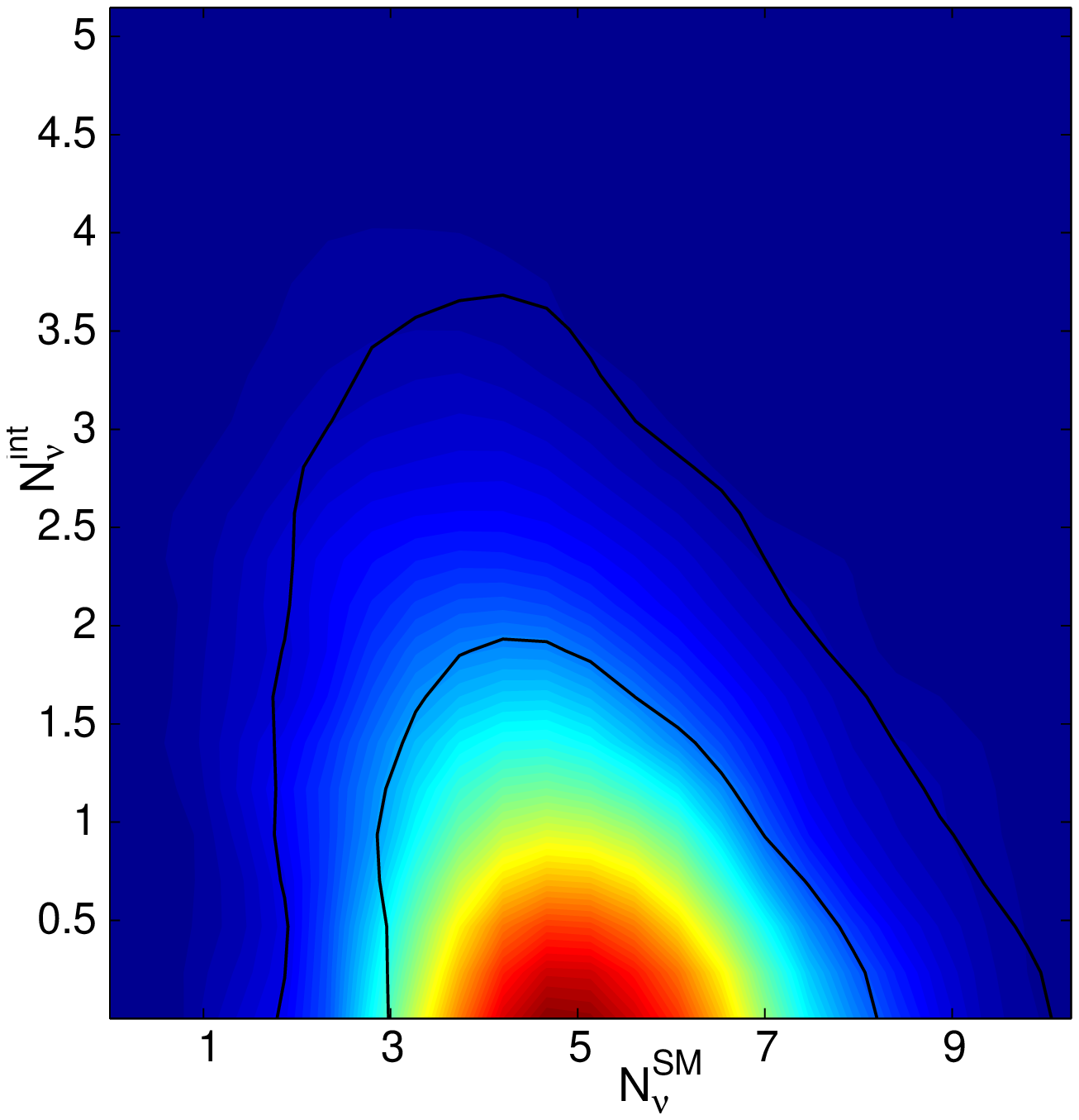}
\includegraphics[width=3.0in]{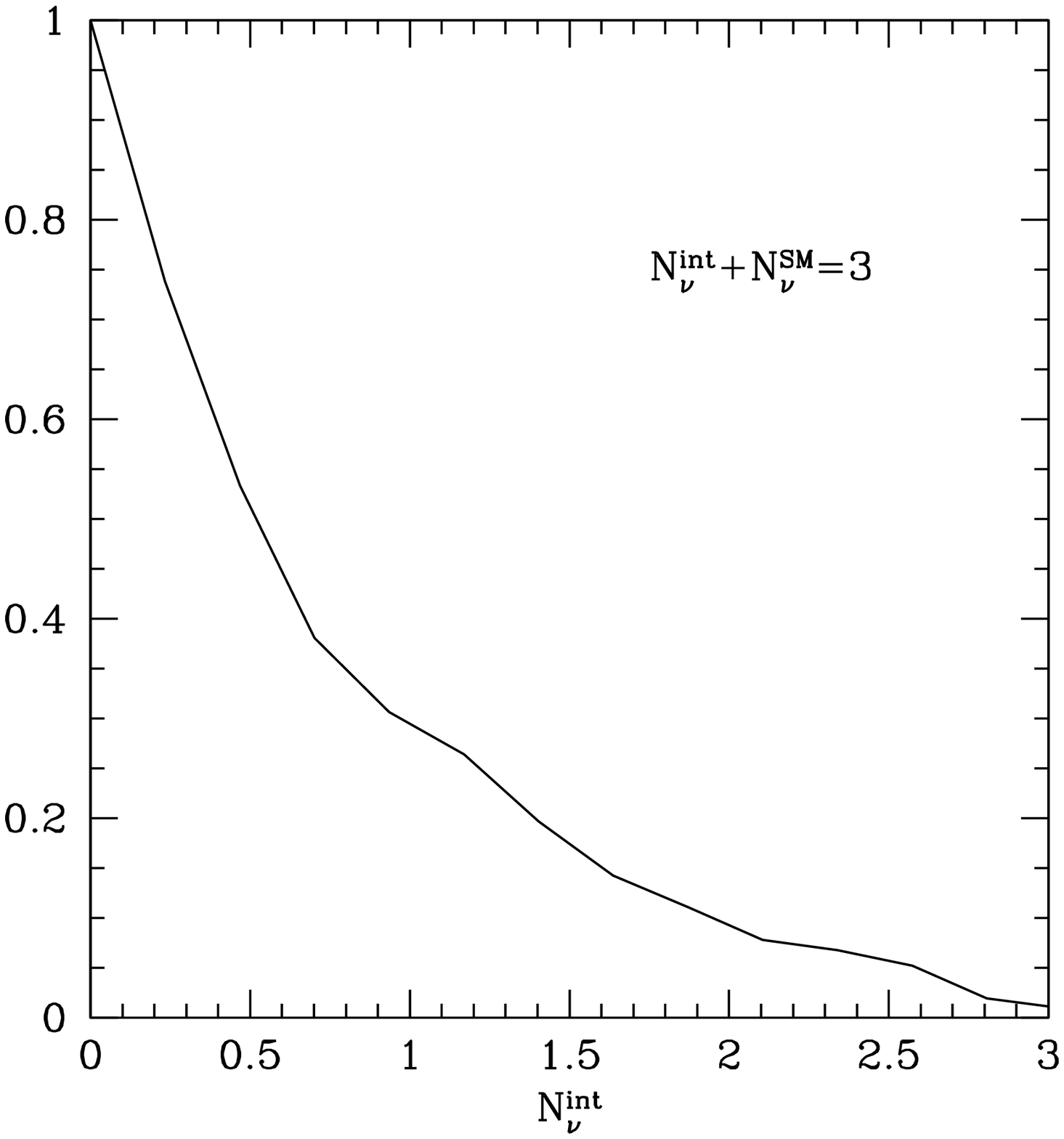}
\caption{Upper: The contour level for the parameters $N_\nu^{\rm SM}$
and $N_\nu^{\rm int}$ in Model A.  The two solid lines are one and two
$\sigma$ contours, shades are from the mean likelihood, using the
\emph{Cosmo} data set.  Lower: The relative likelihood along the line
$N_\nu^{int}+N_\nu^{\rm SM}=3$. }
\label{ninni}
\end{figure} 

It is clear from Figs.~\ref{ninnimarg} and \ref{ninni} that models
with fewer than three interacting neutrinos are favored in a Bayesian
sense.  This does not contradict our previous statement that
cosmological parameter values can be found with $\Nnuint \simeq 3$
which provide a good global fit to the data considered (a reduced
$\chi^2$ statistically consistent with 1).  It is worth elaborating on
the implications and interpretation of this Bayesian limit and why no
contradiction exists.  Firstly, the likelihoods shown are after
marginalizing over all other parameters assuming flat priors in
$\NnuSM$, $\Nnuint$, and all other cosmological parameters.  As such,
the likelihood in Fig. \ref{ninni} is not straightforward measure of
the goodness-of-fit of models with given values of $(\NnuSM,\Nnuint)$
but also of the \emph{density} of good-fitting models with the values
$(\NnuSM,\Nnuint)$ in the hyperplane of parameter space defined by the
remaining cosmological parameters --- it technically quantifies the
cumulative relative likelihood of two populations of models.  We can
certainly conclude, for instance, from Fig.~\ref{ninni} that (with the
aforementioned flat priors) the set of data we have used prefers the
population of models with $(\NnuSM,\Nnuint)\simeq(3,0)$ relative to
the population of models with $(\NnuSM,\Nnuint)\simeq(0,3)$ at more
than 2-$\sigma$.  However, we can \emph{not} conclude from
Fig.~\ref{ninni} that \emph{all} models with
$(\NnuSM,\Nnuint)\simeq(0,3)$ are ruled out at more than 2-$\sigma$ by
current data.\footnote{The notion of ``more than 2-$\sigma$'' depends
on the context.  In reference to the marginalized likelihood in a two
dimensional parameter plane it refers to a given point in this
parameter plane being outside the 95\% confidence region.  In
reference to a particular model with a given set of cosmological
parameters it refers to that model having a reduced $\chi^2$ more than
two standard deviations (of the $\chi^2$ distribution) away from $1$.}
This conclusion can only be made if \emph{no} models with
$(\NnuSM,\Nnuint)\simeq(0,3)$ can be shown to be consistent with the
data, which as we have discussed, is not the case.  Similarly, while
we can conclude that the population of models with $(\NnuSM,\Nnuint)
\simeq (3,0)$ are disfavored compared to the population of models with
$(\NnuSM,\Nnuint) \simeq (5,0)$ we can not conclude that all models
with $5$ free-streaming neutrinos provide better global fit to the
data than all models with $3$ free-streaming neutrinos.  This is just
the nature of Bayesian inference --- it attempts to quantify the
relative likelihood of models within a given paradigm, but does not
make an absolute judgment on the viability of the paradigm itself (or
in this case a particular subset of the models within the
paradigm). To make a judgment on the viability of a given paradigm, a
better approach is to ask the question ``Are there any models within
this paradigm which are consistent with the data?''.  If the answer to
this question is ``yes'' then that paradigm is still viable.  In this
case, another way of summarizing the state of affairs is that even if
priors stipulating $(\NnuSM,\Nnuint)\simeq(3,0)$ or
$(\NnuSM,\Nnuint)\simeq(0,3)$ are imposed to restrict our paradigm to
either the standard model of particle physics or an alternative model
with interacting neutrinos, then models that are globally consistent
with the data can still be found (albeit fewer of them in the later
case).

Finally, let us comment on parameter degeneracy.  Because the number
of neutrinos contribute to setting the redshift of equivalence, both
$\NnuSM$ and $\Nnuint$ are degenerate with $\Omega_m$ and $H_0$.  It
has been pointed out \cite{Pierpaoli, Nnu} that a higher number of
massless neutrinos can be compensated by a higher $H_0$.  We find that
even the marginalized likelihood for $H_0$ in this model is around 80
${\rm km}\ {\rm s}^{-1}\ {\rm Mpc}^{-1}$, which is only one $\sigma$ away from
the HST quoted best fit.  In this respect, our results are discrepant
with those in~\cite{hannestad}, where much larger values of $H_0$ were
obtained.

\subsection{ Model B: Annihilating neutrinos}

In Fig.~\ref{nam} we plot the marginalized likelihood for the neutrino
mass in model B1 (one interacting neutrino plus 2 standard model
neutrinos) using the \emph{Cosmo} and \emph{Cosmo} + Lyman--$\alpha$
data sets (\emph{Cosmo}Ly$\alpha$).  The curves for B2 are similar to
B1.  In Fig~\ref{nbm}, we show the corresponding likelihood for the
case where all three neutrinos annihilate (model B3).  The best fit
values and the 95 \% C.L.  of the marginalized likelihoods are
reported in table \ref{tab:mlim}, for the model in which either one,
two or three neutrinos interact and annihilate.  
For models B1 and B2 the best fits are at $m_\nu \simeq 0~\rm{eV}$.
Note that our approximation does not allow us to explore $m_\nu \simeq
0~\rm{eV}$, as the assumption that all three neutrino masses are equal
breaks down when $m_\nu \alt \sqrt{\delta m^2_{\rm atm}} \sim
0.05~\rm{eV}$.  For model B3 the best fit is non-zero, and is
discussed further below.

\begin{figure}[t]
\includegraphics[width=3.0in]{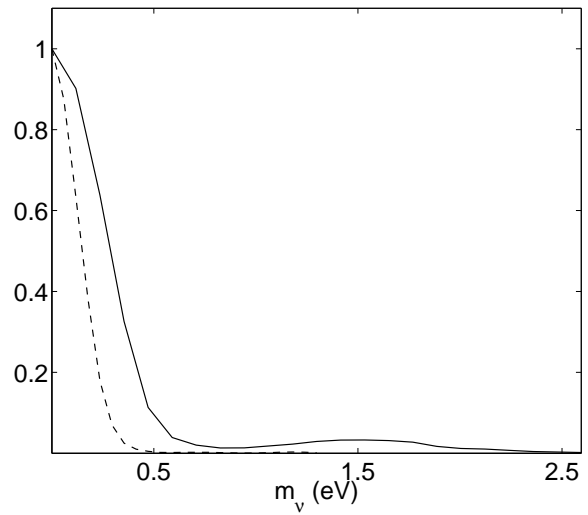}
\caption{Marginalized likelihood for the neutrino mass in Model
B1.  The solid curve and dashed curves correspond to the \emph{Cosmo}
and \emph{Cosmo}Ly$\alpha$ data sets respectively.}
\label{nam}
\end{figure}
\begin{figure}[t]
\includegraphics[width=3.0in]{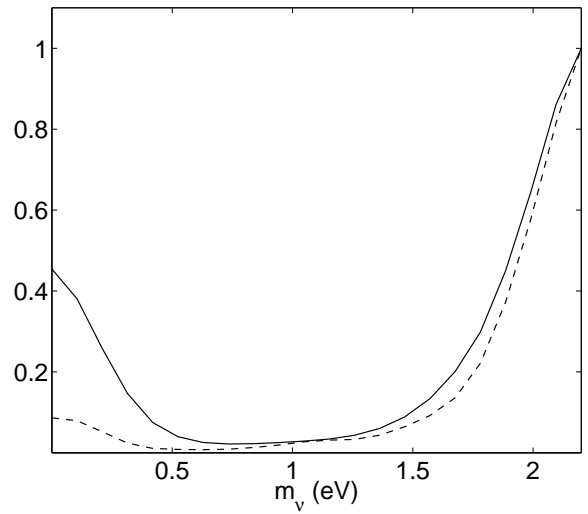}
\caption{Marginalized likelihood for the neutrino mass in the
``neutrinoless universe'' model, B3.  The solid and dashed curves
correspond to the \emph{Cosmo} and \emph{Cosmo}Ly$\alpha$ data sets
respectively.  Note that we imposed a 2.2 eV cut-off for $m_\nu$.}
\label{nbm}
\end{figure}

\begin{table}[h]
      \begin{tabular}{c||c|c||c|c||c|c} 
\hline\hline
 & B1 & B1  & B2  & B2 & B3 & B3 \\
Data set & best fit  & limit & best fit  & limit  & best fit & limit \\
\hline\hline
   \emph{Cosmo} & $<0.1$  &  1.5  &  $<0.1$ &  0.42  & 2.2  & 2.2  \\
\emph{Cosmo}Ly$\alpha$ & $<0.1$ &  0.24  &  $<0.1$ &  0.31  & 2.2  &  2.2 \\
\hline\hline
   \end{tabular}
   \caption{The limits on the neutrino mass, in the various
models. All values are in eV.  For each model, the upper limit refers
to the 95 \% C.L. limit of the marginalized likelihood.  For
$m_\nu<0.1$ eV, the approximation that all three neutrino masses are
degenerate is not satisfied.  Note that for model B3, we imposed a 2.2
eV cut-off, as implied by the tritium beta decay neutrino mass
limit. }
   \label{tab:mlim}
\end{table}

For the cases B1 and B2 where only one or two neutrino species
annihilate, the remaining neutrinos contribute to the dark matter
density today.  As expected, Lyman-$\alpha$ data significantly
tightens the neutrino mass limit in these models, because it sets a
constraint on the overall normalization and the shape of the power
spectrum.  This reduces the 95 \% C.L.  upper limit by quite a lot,
bringing it to 0.24 eV for one interactive neutrino (model B1) and
0.31 eV for 2 interactive neutrinos (model B2).  The addition of the
CBI polarization data does not improve the constraints on neutrinos
mass with respect to those obtained with the \emph{Cosmo} data.  This
is because such low neutrino masses don't have a significant impact on
the CMB power spectrum (see Fig.~\ref{fig:na1_cl}).  As for
degeneracies, the neutrino mass parameter is degenerate with
$\Omega_m$ and $\Omega_\Lambda$ and $H_0$, all of which affect the
redshift of equivalence and therefore the amplitude of the matter
power spectrum at small scales.  As a consequence, $m_\nu$ is also
degenerate with the amplitude of the matter power spectrum $\sigma_8$.
The mean of the marginalized $H_0$ likelihood plot is 68 ${\rm km}\
{\rm s}^{-1}\ {\rm Mpc}^{-1}$ and 71 ${\rm km}\ {\rm s}^{-1}\ {\rm
Mpc}^{-1}$ for models B1 and B2, which are perfectly normal values.

The case B3 is different, because here all three neutrinos species
annihilate, leaving no neutrino contribution to the dark matter today.
The effects on the power spectrum are thus milder (see
Fig.~\ref{fig:nb3_cl}), and hence the limits on the neutrino mass are
expected to be significantly weaker.
In particular, the more massive the neutrino is, the earlier it
annihilates (see Fig.~\ref{fig:Neff}).  If the neutrinos annihilate
early enough that all cosmologically interesting scales are outside
the horizon, then models corresponding to different masses only differ
by a slightly different expansion history very early on.  Therefore,
the data cannot distinguish between different high values of $m_\nu$,
as the corresponding power spectra are very similar (see
Fig.~\ref{fig:nb3_cl} and Fig.~\ref{nb2D}.)

\begin{figure}[t]
\includegraphics[width=3.35in]{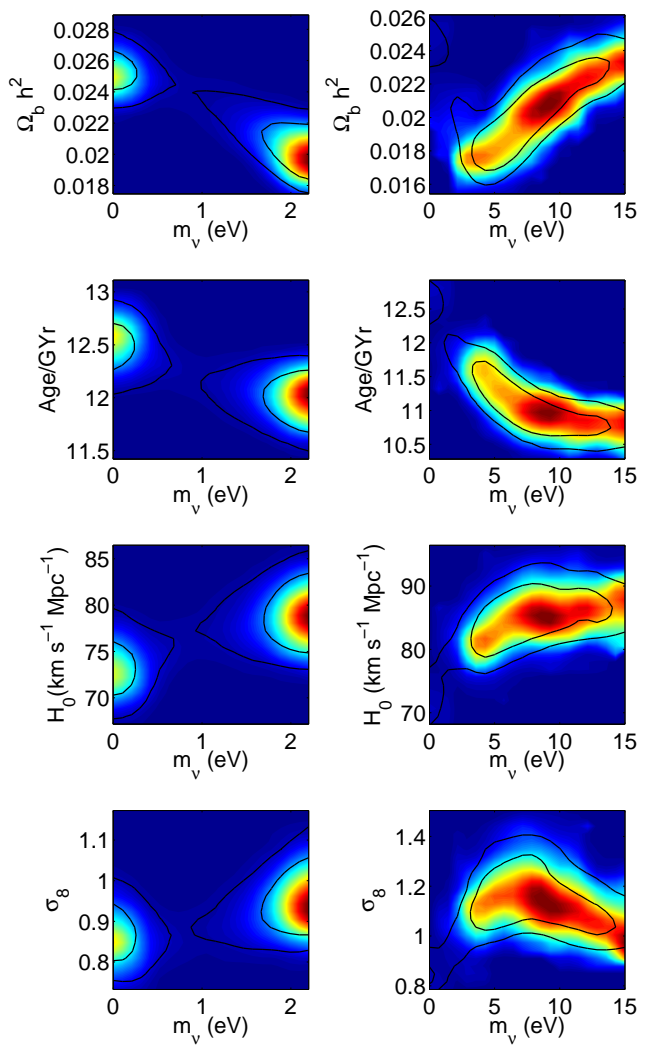}
\caption{The confidence level contours for the parameters $\Omega_b
h^2$, Age, $H_0$, and $\sigma_8$, vs. $m_\nu$, for model B3. The solid lines are the
one and two sigma contours, and shades are from the mean likelihood,
using the {\it Cosmo} data set.  The panels on the left have a 2.2 eV
cut-off imposed on $m_\nu$, while the panels on the right have a 15 eV
cut-off.}
\label{nb2D}
\end{figure}

However, for neutrino masses $m_\nu \simeq 1$ eV, the annihilation
takes place very close to the time of CMB decoupling.  It is during
the annihilation period that the greatest deviation of the sound speed
and the equation of state occur (see Fig.~\ref{fig:w}.)  Hence for
$m_\nu \simeq 1$ eV, the CMB spectrum is affected by the modified
values of $w$ and $c_{s}^2$, in addition, of course, to the lack of
free-streaming.  This tends to disfavor $m_\nu \simeq 1$ eV, with
respect to both larger and smaller values of $m_\nu$, as shown in
Fig.~\ref{nbm}.

In addition, the neutrino mass in model B3 shows significant
degeneracies with the Age and $H_0$, as changes in these parameters
compensate for the effect of increased radiation density on the epoch
of equivalence.  The neutrino mass is also degenerate with $\Omega_b
h^2$ and $\sigma_8$.  We have plotted these degeneracies in
Fig.~\ref{nb2D}.  To help explore the degeneracies, we have determined
the marginalized likelihood contours with and without imposing the 2.2
eV cut-off in $m_\nu$.  We can see from Fig.~\ref{nb2D} that although
a slightly better fit is obtained for larger $m_\nu$ (5-10 eV), this
would imply an unacceptably low Age \cite{Age}, an unacceptably high
$H_0$, and also a value of $\sigma_8 \agt 1$, which is disfavored by
recent cluster number counts~\cite{clusters} and lensing
analysis~\cite{lensing}.  (The baryon abundance however, more closely
matches the BBN determination, $\Omega_b h^2 = 0.022 \pm
0.002$~\cite{BBN}, for larger values of $m_\nu$.)  Imposing the 2.2 eV
cut-off for $m_\nu$ (as required to be consistent with the tritium
beta decay neutrino mass limit) brings these parameters back to more
reasonable values.
However, even in the 0-2.2 eV region, the values of these parameters
seem somewhat discrepant.  For example, the best fit point (at $m_\nu$
= 2.2 eV) has values of $\Omega_b h^2$, Age and $H_0$ which all
deviate from the central values preferred by BBN, globular cluster, and
HST measurements, respectively, by about one sigma.

\subsection{Discussion}

The results above should be compared with the ones obtained by other
authors in the standard case of three massive, non--interacting
neutrinos.  The WMAP team found $m_{\nu} \le 0.23~{\rm eV}$ at 95
\% CL~\cite{Sper03}, using CMB data and the 2dF power spectrum, plus an
assumption on the power spectrum bias which drives most of the
result~\cite{pierpaproc,ElgLah}.  The addition of quite tight
constraints from the SDSS power spectrum leads to an upper limit on
the sum of the three masses of $\sum m_\nu = 0.42~{\rm eV}$ at 95 \%
CL~\cite{selj04}.
The recent results from the
Boomerang team are $ m_\nu < 1~{\rm eV}$ at 95 \% CL from CMB data
alone, and $ m_\nu < 0.4~{\rm eV} $ when the matter power spectrum
data are considered, with no bias assumptions \cite{boom05}.  
Perhaps the most directly comparable value is the
one obtained by \cite{vielproc}, $\sum m_\nu=0.33 \pm 0.27~{\rm eV}$,
with the same Lyman--$\alpha$ sample and comparable assumptions about the
parameters as in our analysis. 
These results are broadly similar to what we find for the cases B1 and
B2, in which one or two standard model neutrinos remain.  As a general
result, we can conclude that the neutrino mass is expected to be below
about 0.5 eV in those cases, and is not significantly affected by the
presence of the tightly-coupled component of the relativistic energy
density.  However, the ``neutrinoless'' model (B3) is different, and
leads to much weaker $m_\nu$ constraints, albeit at the expense of somewhat 
discrepant values for $\Omega_b h^2$, Age and $H_0$.

Finally, we comment on the implications of these results for neutrino
decay.  One consequence of strong $\nu-\phi$ interactions is that
neutrinos may be unstable.  For the range of couplings considered
here, $g \agt 10^{-5}$, the neutrinos could decay over astronomical
distances, which is testable in neutrino telescope
experiments~\cite{decay}.  For small masses, the cosmological
consequences of these couplings are represented by Model A, while for
larger masses they are represented by either Model B2 or B3 (depending 
on whether one or two neutrino species are unstable.)

Ref.~\cite{hannestadraffelt} claims stringent limits on the couplings
$g_{ij}$, and thus on neutrino lifetimes, based upon the fact that
Refs.~\cite{hannestad,trotta} find some evidence for neutrino
free-streaming in the CMB.  However, \cite{hannestad,trotta} analyzed
only a scenario in which all three neutrinos behave in the same way,
and thus certainly cannot be used to claim that {\it all three}
neutrino species must be free-streaming.\footnote{Neutrino decay
requires only some subset of the couplings to be large (e.g. $g_{21}$
allows for the decay $\nu_2 \rightarrow \nu_1 + \phi$) implying that
only a subset of the neutrinos, $\nu_i$, need be interacting.  Without
demonstrating that all the couplings $g_{ij}$ must simultaneously be
small, it is not possible to set stringent cosmological limits on any
particular individual element of $g_{ij}$.}

Given the results here, we conclude that current cosmological data
{\it does not} impose limits on either the individual
couplings, $g_{ij}$, or upon neutrino lifetimes.


\section{Conclusions}

We have investigated the distinction between a free-streaming or
interacting fluid of neutrinos, and explored the constraints on these
neutrinos imposed by cosmology.  Although some of our results are
applicable in a wider context, we have used an example involving
additional couplings between the neutrinos and a light boson.  Our
main conclusion is that {\it models with interacting neutrinos remain
viable}, contrary to the claim in Ref.\cite{hannestad,hannestadraffelt}.

As a general result, we find that both CMB polarization data and
Lyman--$\alpha$ data help to constrain the number of standard model
neutrino species.  The inclusion of Lyman--$\alpha$ data also helps
improve the constraints on neutrino mass.  With the data considered,
we have found upper limits on the neutrino mass, but no detection
of a nonzero neutrino mass.

Two parameterizations of the interacting-neutrino models have been examined:  

(1) In the first, we allowed for an arbitrary number of free-streaming
(standard model) neutrinos, $\NnuSM$, and tightly-coupled
(interacting) neutrinos, $\Nnuint$.  Within the context of the
neutrino-scalar model, this corresponds to the limit $m_{\nu}
\rightarrow 0$ and $m_{\phi} \rightarrow 0$.
We have found that within the ($\NnuSM$,$\Nnuint$) plane, the data
favors free-streaming neutrinos over tightly-coupled neutrinos in a
Bayesian analysis.  However, we find even if the prior $\NnuSM \simeq 0$ is
imposed (so that all neutrinos are interacting) models can be found
that are a good global fit to the data considered, without resorting
to extreme values for the cosmological parameters. In this respect,
models with interacting neutrinos remain viable.  We emphasize that
the constraints obtained here are very general, as $\NnuSM$ and
$\Nnuint$ can parametrize any free-streaming or tightly coupled
relativistic degrees of freedom, which need not consist of neutrinos.

(2) Our second parametrization consists of models with non-zero $m_\nu$.
We fixed the total number of neutrinos to three, and allowed either
one, two, or all three, to interact with a massless boson.  In
this scenario, the interacting neutrinos annihilate to bosons when $T
\sim m_\nu$, thus removing them from the plasma.
In the case where either one or two neutrino species
interact/annihilate, we find the upper limits on neutrino mass are
broadly similar to those for the standard scenario.  However, if all
three neutrino species annihilate to leave a ``neutrinoless
universe'', the neutrino mass limits are significantly weaker.  
In this case, values of $m_\nu$ comparable to the tritium beta
decay limit of 2.2 eV are permitted, although a low Age and a high
$H_0$ tend to somewhat disfavor the scenario.

\begin{acknowledgments}
We thank Matteo Viel for useful discussions and assistance in
implementing the Lyman--$\alpha$ data.  We thank John Beacom and Scott
Dodelson for useful comments on the manuscript.  This work was
supported in part at Caltech by NASA Grant No. NAG5-9821 and DoE Grant No.
DE-FG03-92-ER40701.  EP is supported through the ADVANCE Fellowship
Program (NSF Grant No. AST-0340648), and also by NASA Grant No. NAG5-11489.
KS acknowledges support at Caltech from a Canadian NSERC Postgraduate Scholarship and support
at the IAS from NASA through Hubble Fellowship Grant No. HST-HF-01191.01-A
awarded by the Space Telescope Science Institute, which is operated by
the Association of Universities for Research in Astronomy, Inc., for
NASA, under Contract No. NAS 5-26555.  NFB was supported by the Sherman
Fairchild Foundation at Caltech.
\end{acknowledgments}

\begin{appendix}
\section{Neutrino-Scalar Fluid Properties}
\label{App:A}

Here we summarize a few basic relationships for Fermi-Dirac 
and Bose-Einstein statistics and derive from them the 
quantities $w$, $c_{s}$, and $\Nnuint$ used in Eqs.~(\ref{eqn:perturbations})--(\ref{eqn:energydensity}) to evolve the perturbation dynamics and shown in Fig.~{\ref{fig:Neff}} and Fig.~{\ref{fig:w}}.  We note that these quantities may also be found by solving the continuity equations for energy and entropy density numerically, but we believe the analytical forms we present here are physically illustrative and may be useful in other contexts.

The number density $n_{i}$, energy density $\rho_{i}$, and pressure $P_{i}$ of a fermionic ($\xi=1$) or bosonic ($\xi=-1$) species $i$ of mass $m_{i}$ with $g_{i}$ internal degrees of freedom in thermal equilibrium can be written as 
\begin{align}
n_{i} &=\frac{g_{i}}{2 \pi^2} \int_0^{\infty} p^2 dp \frac{1}{1+\xi e^{E/T}} \nonumber \\
&= \frac{g_{i} m_{i}^3}{2\pi^2}\sum_{n=1}^{\infty}\frac{(-\xi)^{\scriptscriptstyle n+1}}{\rn x}\left[K_2(\rn x)\right] \nonumber \\
&\rightarrow g_{i}\left[\frac{3}{4}+\frac{1}{8}(1-\xi)\right]\frac{\zeta(3)}{\pi^2} T^3 \\
\rho_{i} &=\frac{g_{i}}{2 \pi^2} \int_0^{\infty} p^2 dp \frac{E}{1+\xi e^{E/T}} \nonumber \\
&= \frac{g_{i} m_{i}^4}{2\pi^2}\sum_{n=1}^{\infty}\frac{(-\xi)^{\scriptscriptstyle n+1}}{\rn x}\left[K_1(\rn x) + \frac{3}{\rn x}K_2(\rn x)\right] \nonumber \\
&\rightarrow g_{i}\left[\frac{7}{8}+\frac{1}{16}(1-\xi)\right]\frac{\pi^2}{30} T^4 \\
P_{i} &=\frac{g_{i}}{6 \pi^2} \int_0^{\infty} p^2 dp \frac{p^2/E}{1+\xi e^{E/T}} \nonumber \\
&= \frac{g_{i} m_{i}^4}{2\pi^2}\sum_{n=1}^{\infty}\frac{(-\xi)^{\scriptscriptstyle n+1}}{\rn x}\left[\frac{1}{\rn x}K_2(\rn x)\right] \nonumber \\
&\rightarrow g_{i}\left[\frac{7}{8}+\frac{1}{16}(1-\xi)\right]\frac{\pi^2}{90} T^4 
\end{align}
where $E=\sqrt{p^2+m_{i}^2}$, $x \equiv m_{i}/T$, $K_\alpha$ are modified Bessel functions of the second kind, $\zeta(3)=1.2020569$ is the Riemann zeta function of three, and the arrows indicate the high-temperature limit $x \rightarrow 0$.  Notice that the standard result for massless particles $P_{i}=\rho_{i}/3$ is recovered in this limit.  These results can be straightforwardly derived by expanding the distribution function as a geometric series of Boltzmann factors.  This form is useful for tabulating the density and pressure as the $K_\alpha$ can be rapidly evaluated and the sum converges quickly and can be truncated to the desired accuracy --- keeping only the leading term is the Maxwell-Boltzmann approximation to the distribution function.

Specializing to the fermionic case ($\xi=1$) we write the energy density and pressure of a single fermionic degree of freedom as
\begin{align}
\rho_{f} &\equiv \frac{7}{8}\frac{\pi^2}{30} \chi_{\rho}(x)T^4 \nonumber \\
&= \frac{m_{f}^4}{2\pi^2}\sum_{n=1}^{\infty}\frac{(-)^{\scriptscriptstyle n+1}}{\rn x}\left[K_1(\rn x)  + \frac{3}{\rn x}K_2(\rn x)\right] \\
P_{f} &\equiv \frac{7}{8}\frac{\pi^2}{90} \chi_{P}(x)T^4 \nonumber \\
&= \frac{m_{f}^4}{2\pi^2}\sum_{n=1}^{\infty}\frac{(-)^{\scriptscriptstyle n+1}}{\rn x}\left[\frac{1}{\rn x}K_2(\rn x)\right] \, ,
\end{align}
where $\chi_{\rho/P} \rightarrow 1$ for $x \rightarrow 0$ but are $< 1$ for finite $x$.  We also find it useful to define
\begin{align}
\frac{d\rho_{f}}{dT} &\equiv 4 \cdot \frac{7}{8}\frac{\pi^2}{30} \chi_{d\rho}(x)T^3 \nonumber \\
&= \frac{4}{T}\cdot \frac{m_{f}^4}{2\pi^2}\sum_{n=1}^{\infty}\frac{(-)^{{\scriptscriptstyle n+1}}}{\rn x}\left[\frac{3}{4}K_1(\rn x)  + \frac{12+\rn x}{4 \rn x}K_2(\rn x)\right] \\
\frac{dP_{f}}{dT} &\equiv 4 \cdot \frac{7}{8}\frac{\pi^2}{90} \chi_{dP}(x)T^3 \nonumber \\
&= \frac{4}{T}\cdot \frac{m_{f}^4}{2\pi^2}\sum_{n=1}^{\infty}\frac{(-)^{{\scriptscriptstyle n+1}}}{\rn x}\left[\frac{1}{4}K_1(\rn x)  + \frac{1}{ \rn x}K_2(\rn x)\right] \, ,
\end{align}
where similarly $\chi_{d\rho/dP} \rightarrow 1$ for $x \rightarrow 0$.

Let us now consider, as we do in this paper, a thermalized fluid at temperature $T_{\nu\phi}$ consisting of $N$ massive neutrinos coupled to a single massless scalar degree of freedom $\phi$.  

The total energy density and pressure of this fluid are
\begin{align}
\rho = \rho_{\phi}+\rho_{\nu}=\frac{\pi^2}{30}T_{\nu\phi}^4 + 2 N\frac{7}{8}\frac{\pi^2}{30}\chi_{\rho}(x_{\nu})T_{\nu\phi}^4
\label{eqn:rhorho}
\end{align}
and
\begin{align}
P = P_{\phi}+P_{\nu}=\frac{\pi^2}{90}T_{\nu\phi}^4 + 2 N\frac{7}{8}\frac{\pi^2}{90}\chi_{P}(x_{\nu})T_{\nu\phi}^4
\end{align}
where the factor of $2$ in the neutrino terms accounts for anti-neutrinos and $x_{\nu} \equiv m_{\nu}/T_{\nu\phi}$.  

We can now write the equation of state $w=P/\rho$ in terms of $\chi_{\rho}$ and $\chi_{P}$ as
\begin{align}
w = \frac{1}{3}\left[\frac{4/7 + N \chi_{P}(x_{\nu})}{4/7+ N \chi_{\rho}(x_{\nu})}\right] \, .
\label{eqn:wform}
\end{align}

Similarly, we can write the soundspeed $c_{s}^2=dP/d\rho$ in terms of $\chi_{d\rho}$ and $\chi_{dP}$ as
\begin{align}
c_s^2 = \frac{1}{3}\left[\frac{4/7 + N \chi_{dP}(x_{\nu})}{4/7 + N \chi_{d\rho}(x_{\nu})}\right] \, .
\label{eqn:cs2form}
\end{align}

The form of Eq.~(\ref{eqn:wform}) and Eq.~(\ref{eqn:cs2form}) explains the behavior shown in Fig.~\ref{fig:w}.  For $T_{\nu\phi} \gg m_{\nu}$ we have $\chi \rightarrow 1$ and both $w$ and $c_{s}^2$ approach $1/3$. For $T_{\nu\phi} \ll m_{\nu}$ we have $\chi \rightarrow 0$ and again $w$ and $c_{s}^2$ approach $1/3$.  It is only during the annihilation of the neutrinos ($T_{\nu\phi} \sim m_{\nu}$) that the values of $w$ and $c_{s}^2$ deviate from those for a relativistic fluid. Larger values for $N$ result in larger deviations.

We now derive an expression for the temperature of the neutrino-scalar fluid $T_{\nu\phi}$ as a function of time (measured with $T_{\gamma}$).  In the standard case the neutrino temperature just falls as $\TnuSM =(4/11)^{1/3} T_\gamma \propto a$.  It is convenient to measure $T_{\nu\phi}$ relative to this standard case and define the ratio ${\cal R}_{\nu\phi} \equiv T_{\nu\phi}/\TnuSM$. For the times of interest the weak interactions have decoupled already and the comoving entropy density $S = a^3(\rho + P)/T_{\nu\phi}$ of the neutrino-scalar fluid is constant.  The constancy of $S$ implies that
\begin{align}
{\cal R_{\nu\phi}} = \left(\frac{4/7 + N}{4/7 + N[(3/4)\chi_{\rho}(x_{\nu})+(1/4)\chi_{P}(x_{\nu})]}\right)^{1/3} \, ,
\end{align}
which, recalling that $x_{\nu}=m_{\nu}/(R_{\nu\phi}\TnuSM)$, is a transcendental equation which implicitly determines ${\cal R}_{\nu\phi}$ as a function of $\TnuSM$.

Now if we write the energy density in terms of an effective number of standard model neutrinos
\begin{align}
\rho = \Nnuint \frac{7}{8}\frac{\pi^2}{15}\left(\TnuSM\right)^4
\end{align}
we find, comparing with Eq.~(\ref{eqn:rhorho}), that
\begin{align}
\Nnuint &= \left[\frac{4}{7} + N \chi_{\rho}(x_{\nu})\right]{\cal R}_{\nu\phi}^4 \nonumber \\
&\rightarrow \frac{4}{7}\left[1+\frac{7}{4}N\right]^{4/3} \, ,
\end{align}
where the last line holds for $T_{\nu\phi} \ll m_{\nu}$ (after annihilation is complete). In accordance with Fig.~\ref{Neff} we find that $\Nnuint \rightarrow (2.20,4.25,6.58)$ for $N=(1,2,3)$ respectively.          
\end{appendix}


\end{document}